\documentclass[12pt]{article}

\usepackage[margin=1in]{geometry}
\usepackage[utf8]{inputenc}
\usepackage[T1]{fontenc}
\usepackage{amsmath, amssymb, amsthm}
\usepackage{graphicx}
\usepackage{booktabs}          
\usepackage{hyperref}
\usepackage{microtype}
\usepackage{lineno}            
\usepackage{setspace}
\usepackage{footnote}
\usepackage{comment}
\usepackage{subcaption}
\usepackage[numbers, compress, sort, super]{natbib}
\usepackage{pdflscape}
\usepackage{listings}
\usepackage{xcolor}
\usepackage{relsize}

\newtheorem{ass}{Assumption}

\newtheorem{algo}{Algorithm}

\lstdefinestyle{simTutorialR}{
  language=R,
  basicstyle=\ttfamily,
  numbers=none,
  showstringspaces=false,
  breaklines=true,
  frame=none,
  keywordstyle=\color{black},   
  commentstyle=\color{gray!70},
  stringstyle=\color{black},    
  columns=fullflexible
}

\newcommand{\SPCE}{\mbox{\tiny{SPCE}}}

\newcommand{\mP}{\mathbb{P}}
\newcommand{\mr}{\mathrm{mr}}
\newcommand{\Var}{\mathrm{Var}}

\DeclareMathOperator\bE{\mathbb E} 

\DeclareMathOperator\logit{\mathrm logit}

\def\df{\emph{10}}
\def\nt{\emph{00}}
\def\at{\emph{11}}
\def\co{\emph{01}}

\title{Implementing the principal stratum strategy for intercurrent events with survival outcomes: a tutorial}
\author{
  Xiaoxiao Zhou$^{1}$ \and
  Joyce Chen$^{2}$ \and
  Pallavi Mishra-Kalyani$^{2}$ \and
  Xiaoxue Li$^{2}$ \and
  Yuan Li Shen$^{2}$ \and
  Shu Wang$^{2}$ \and
  Susan Halabi$^{3}$ \and
  Fan Li$^{4,*}$
  \\[1em]
  \small $^1$Department of Biostatistics, University of Alabama at Birmingham, Alabama, USA \\
  \small $^2$Office of Biostatistics, CDER, U.S. Food and Drug Administration, Maryland, USA \\
  \small $^3$Department of Biostatistics and Bioinformatics, Duke University, North Carolina, USA \\
  \small $^4$Department of Statistical Science, Duke University, North Carolina, USA \\[0.5em]
  \small $^*$Corresponding author: Fan Li, Department of Statistical Science, Duke University\\
  \small \href{mailto:fl35@duke.edu}{fl35@duke.edu}
}

\date{} 

\begin{document}

\maketitle

\begin{abstract}
The International Council for Harmonization (ICH) E9 (R1) addendum provides the estimand framework to formulate treatment effects in a clinical trial. One of the attributes of an estimand the framework describes is intercurrent events. Among the five strategies to intercurrent events the guidance lists, the principal stratum strategy is the most conceptually and technically challenging because it defines treatment effects on unobserved strata. Its application to survival outcomes is particularly inaccessible to practitioners. This tutorial reviews the methodology and implementation of the estimand framework with the principal stratum strategy to address intercurrent events with survival outcomes. We illustrate using a clinical trial in oncology and focus on a simple case with binary treatment and a single binary intercurrent event of discontinuation of the assigned treatment. We define the causal effects and review two main methods for estimating the effects: the mixture model method and the weighting method. For each method, we elaborate the associated assumptions, models, sensitivity analysis, software and provide example R code. We conduct simulation studies that mimic the real study to study the operation characteristics of these methods.

\end{abstract}

\section{Introduction}




Intercurrent events (ICEs) are common in clinical trials; these are post-randomization events that can affect either the interpretation or the existence of measurements relevant to the clinical question of interest. There are many types of ICEs in clinical trials and we review two main types. The first type is non-adherence to the assigned treatment, including switching or discontinuation of the treatment. For example, in a trial to compare the efficacy of two prostate cancer treatments, the primary endpoint is the overall survival (OS) time, but a substantial portion of patients discontinued the assigned treatment due to toxicity, which is potentially caused by the treatment and affects the OS time. The second type -- common in oncology trials -- is the intermediate disease or complication status between the initial treatment and the final clinical endpoint. We give three examples. A first example concerns the progression-free survival time (PFS) in a slow-growing tumor trial, where the ICE is the symptomatic progression of the cancer, e.g., tumor growth, new lesions, or worsening disease.  A second example concerns the survival time for patients who had achieved complete response/remission (CR) or partial response (PR) after the initial treatment in a malignant hematology trial, where the ICE is the CR/PR status. A third example concerns the complication-free survival time after allogeneic stem cell or bone marrow transplantation, where the ICE is the post-treatment complication of Graft-versus-Host Disease (GVHD).  Often both types of ICEs are present in a clinical trial. The presence of ICEs in a clinical trial complicates the definition of causal estimands and statistical analysis.  

The ICH E9 (R1) addendum extends the original ICH E9 guidelines to include statistical principles related to estimands and sensitivity analysis in clinical trials. The addendum provides the estimand framework -- a structured approach to discuss the treatment effect(s) of interest that a clinical trial should address and align design, conduct, analysis and interpretation with the clinical trial objectives and questions of interest. The estimand consists of five key attributes: population, treatment, endpoint, population-level summary, and strategies to address ICEs. The addendum lists five strategies to address ICEs: treatment policy, hypothetical, principal stratum, while on treatment, and composite strategies. These can be applied separately or in combination, depending on the nature of the ICEs and the study objectives.

Recent research has extensively discussed the general principles of the estimand framework, focusing on selecting and formulating appropriate estimands. \cite{mallinckrodt2020aligning, lipkovich2020causal, stensrud2022translating, han2023defining, morgan2025applying, beckers2025adopting} However, practical guidance on the implementation of each ICE strategy remains limited. In particular, the principal stratum strategy has been relatively less adopted, partially due to its conceptual and computational complexity. This strategy is based on principal stratification,\cite{Frangakis2002} which is a causal inference framework for defining estimands in the presence of post-treatment confounding factors, e.g., an ICE. Principal stratification is a classification of subjects based on their joint counterfactual ICE statuses under both treatment and control, i.e., the principal strata, and the target estimand is the treatment effects within a principal stratum, the mathematical formulation of which is given in Section \ref{sec:setup}. The choice of target estimands depends on the specific clinical context. For example, when the ICE is treatment discontinuation, a target estimand is the treatment effect in overall survival time for the principal stratum of patients who can tolerate and continue the assigned treatment regardless of the initial assignment. When the ICE is cancer progression, a target estimand is the treatment effect in survival time for the principal stratum of patients whose cancer would not progress regardless of the initial assignment.  


Because each patient is observed under only one treatment, patients within a principal stratum cannot be identified directly, neither before the trial nor using data from a parallel-group randomized trial. Therefore, principal strata are latent subpopulations,  differing from the subpopulations in the per-protocol analyses, where the membership is determined solely by the observed ICE status under the assigned treatment. Additional structural and modeling assumptions are necessary to identify these latent strata and associated causal effects. \citep{zhang2009likelihood,  jiang2022multiply, lipkovich2022using} In the literature, there are two main estimation approaches: (i) the mixture model approach based on the exclusion restriction assumption; \cite{imbens1997bayesian,zhang2009likelihood,Mattei2013} (ii) the principal score weighting approach based on the principal ignorability assumption. \cite{jo2009use, ding2017principal, jiang2022multiply,cheng2024multiply} Implementing either approach requires substantial effort in statistical computation and programming.  
Survival or time-to-event endpoints, prevalent in clinical research, bring additional challenges to implementing the principal stratum strategy. To this end, Liu and Li \cite{liu2023pstrata} developed the mixture-model-based estimation method with the associated \textbf{PStrata} R package, and Cheng et al. \citep{cheng2026multiply} developed the principal-score-weighting estimation method with the associated \textbf{mrPStrata} R package. \cite{mrPStrata} Despite these advances, the gap between theory and practice remains wide: the principal stratum strategy continues to be poorly understood in regulatory studies.  Noirjean et al. \cite{noirjean2025implementation} provided a recent case study that compares the hypothetical strategy with the principal stratum strategy in a vaccine trial, but did not supply detailed information on implementation. This gap motivates us to provide a tutorial for implementing the estimand framework with a principal stratum strategy with survival outcomes. 

We acknowledge that the principal strata strategy is conceptually controversial in the causal inference literature. Some authors caution against causal effects defined in the latent principal strata because they lack transparency and require implausible assumptions to identify and instead advocate mediation analysis, \cite{stensrud2022translating, vansteelandt2025chasing} while other authors \cite{vanderweele2011principal,ding2025roles} adopt a more balanced view, recognizing that principal causal effects can provide refined clinical insights beyond the conventional intention-to-treat effects. Instead of taking a firm position in the ongoing conceptual debate, our goal is to provide potential users with detailed information on the principal stratum strategy so that they could make an informed decision on whether to adopt the strategy in their specific study.

To simplify the discussion, we focus on a single type of ICE--treatment discontinuation--and demonstrate both the mixture modeling and weighting estimation methods. Section \ref{sec:setup} introduces the setup in the case of a single binary ICE and extensions. Section \ref{sec:ER} and \ref{sec:PI} introduce the estimation method and sensitivity analysis of each of the two approaches, respectively. Section \ref{sec:software} discusses software implementation. Section \ref{sec:application} illustrates the methods using an oncology trial. Section \ref{sec:simu} conducts extensive simulation studies based on the real trial to further examine the operating characteristics of the methods.




\section{Review of principal stratification: estimands and assumptions} \label{sec:setup}
\subsection{An illustrating oncology trial with treatment discontinuation}
Our running example is the CALGB 90206 trial, \citep{rini2008bevacizumab,rini2010phase} a prospective, randomized multicenter phase III randomized trial on the efficacy of bevacizumab on the prognosis of patients with metastatic renal cell carcinoma (RCC). RCC has long been recognized as a chemotherapy-refractory malignancy. The immune system’s role in RCC biology has prompted investigation into immunotherapeutic agents such as interferon alfa (IFN), a cytokine with immunomodulatory properties. IFN was historically a first-line standard of care for metastatic RCC, achieving objective response rates (ORR) of $10\%$-$15\%$ and a median overall survival (OS) of approximately 12 months. \citep{MRC1999} Bevacizumab, a monoclonal antibody targeting vascular endothelial growth factor, has also demonstrated activity in metastatic RCC. In the CALGB 90206 trial, the treatment arm received bevacizumab plus IFN and the control arm received the IFN monotherapy.  

Between 2003 and 2005, the trial recruited 732 previously untreated RCC patients with $363$ and $369$ patients in the treatment and control arm, respectively. The primary endpoint was a time-to-event outcome: the overall survival (OS) time, with a median OS of 18.3 months ($95\%$ CI: 16.5 to 22.5 months) in the treatment arm, compared to 17.4 months ($95\%$ CI: 14.4 to 20.0 months) in the control arm. The p-value of the unstratified log-rank test of the OS between the two arms is 0.097, representing a positive but statistically non-significant intention-to-treat (ITT) effect of bevacizumab plus IFN versus the IFN monotherapy. However, the trial reported high rates of treatment discontinuation due to toxicity, disease progression, and other factors that are typical in oncology trials. After discontinuing the assigned treatment, some patients may switch to the opposite arm or to second-line therapies, e.g. older agents or other agents as directed by the physician.
These ICEs complicate the interpretation of the primary outcome. In this paper, we re-analyze the trial using the principal stratification strategy to address treatment discontinuation. 
To simplify the illustration, we will group all types of discontinuation into a single category in the analysis, but acknowledge the limitation of this. A detailed discussion is given in Section 5.1. 

\subsection{Basic formulation with a single binary intercurrent event} \label{sec:basic-setup}
We use potential outcomes to formulate principal stratification. Consider a two-arm randomized controlled trial with $N$ subjects. For each participant $i$, let $Z_i \in \{0, 1\}$ denote the randomized assigned treatment, where $Z_i = 1$ indicates active treatment and $Z_i = 0$ indicates control, corresponding to the bevacizumab plus IFN and the IFN monotherapy arm in the CALGB 90206 trial, respectively. Let $X_i$ be the set of baseline covariates, $C_i$ the observed censoring time, and $T_i$ the time-to-event outcome, e.g. the overall survival time. Each subject has two potential failure time $T_i(z)$ and censoring time $C_i(z)$, for $z= 0, 1$, respectively. The \emph{potential} survival distribution under assignment $z$ is $S_{z}(t) = \Pr(T_i(z) > t)$. A standard causal estimand is the ITT survival probability causal effect:\cite{mao2018propensity}
\begin{equation}
    \tau^{\SPCE}(t) = \bE[S_{1}(t)] - \bE[S_{0}(t)], \quad 0 \leq t \leq t_{\max},
\end{equation}
where $t_{\max}$ is the maximum follow-up time. Other estimands include the marginal hazard ratio and the restricted average causal effect.\cite{liu2024principal} An ICE, denoted by $D_i$, occurs after the treatment initiation and affects either the interpretation or the existence of the outcome. Because $D_i$ is measured after randomization, it has two potential values $D_i(z)$ for $z=0,1$. 
In the CALGB 90206 trial, $D_i$ is the indicator of treatment discontinuation, with $D_i=1$ denoting that patient $i$ discontinued the assigned treatment and $D_i=0$ otherwise. Consequently, $D_i(z)$ is the potential status of treatment discontinuation if the patient $i$ were assigned to arm $z$.
In the presence of ICE, we extend the potential outcomes notation to 
$C_i(z, D_i(z))$ and $T_i(z, D_i(z))$, as the potential censoring time and failure time if the patient $i$ were assigned to arm $z$ and experienced the ICE $D_i(z)$, respectively. Among these potential outcomes, only those corresponding to the actual assigned arm $Z_i$ are observed, with $D_i = D_i(Z_i)$, $C_i = C_i(Z_i, D_i(Z_i))$, $T_i = T_i(Z_i, D_i(Z_i))$, and the other potential outcomes are missing.

A principal stratification with respect to the ICE, $D$, is the classification of patients based on the joint potential values of $D$, namely, the principal stratum, $U_i = (D_i(0), D_i(1))$. \citep{Angrist1996, Frangakis2002} With binary $Z$ and $D$, there are four strata: (i) $U_i = (1, 1) = \at$: participants who would discontinue the assigned treatment regardless of their initial assignment; (ii) $U_i = (0, 1) = \co$:  participants who would discontinue the treatment if assigned to the active treatment arm but would continue if assigned to the control arm; 
(iii) $U_i = (1, 0) = \df$: participants who would discontinue the treatment if assigned to the control arm but would continue if assigned to the active treatment arm; (iv) $U_i = (0, 0) = \nt$: participants who would continue the assigned treatment regardless of their initial assignment.
When the primary ICE is treatment discontinuation, arguably only the $\nt$ stratum is informative of the treatment efficacy, and thus it will be our target population, which we label as \emph{always-continuer} to emphasize its practical meaning. Similarly, we will label the $\at$ stratum as \emph{always-discontinuer}. 

The principal stratum is, by definition, unaffected by the treatment assignment and thus can function as a (latent) pre-treatment variable. This allows comparisons of $S_1(t)$ and $S_0(t)$ within a principal stratum to have a causal interpretation. Denoting the potential survival distribution under assignment $z$ for stratum $u$ as $S_{z,u}(t) = \Pr(T_i(z) > t \mid U_i = u)$, we define the survival principal causal effects (SPCE) estimand for stratum $u$ as 
  \begin{equation}
	\begin{aligned}
	      \tau_u^{\SPCE} (t)  & = \bE[S_1(t) - S_0(t) \mid U_i = u] = \bE[S_{1,u}(t)] - \bE[S_{0,u}(t) ], \quad \text{for all } u.
	\end{aligned}
\end{equation}
In contrast, directly stratifying on the observed $D_i$ when comparing the outcomes, namely, $E\{ S_1(t)  \mid D_i = d, Z_i=1 \} - \bE\{  S_0(t) \mid D_i = d , Z_i=0\}$, does not have a causal interpretation, \citep{Rosenbaum1984} because the groups $\{ i: D_i = d, Z_i=1 \}$  and $\{ i: D_i = d, Z_i=0 \}$ are not the same cohorts of subjects when $Z$ affects $D$. The standard ITT is a weighted sum of the SPCE: 
	$
	\tau^{\SPCE}(t) = \sum_{u} \pi_u \tau^{\SPCE}_u(t),
	$
    where $\pi_u  = \Pr( U_i=u) $ is the proportion of the stratum $u$. This shows that principal stratification characterizes a type of treatment effect heterogeneity between different subpopulations that may not be captured by observed covariates. 

\begin{table}[ht]
    \centering
    \renewcommand{\arraystretch}{1.2} 
    \caption{Notation table} 
    \label{table:notations}
    \resizebox{\textwidth}{!}{%
    \begin{tabular}{ll} 
        \hline
        \textbf{Notation} & \textbf{Description} \\ 
        \hline
        $Z_i$ & (binary) treatment assignment indicator for individual $i$ \\ 
        $D_i$ & (binary) intercurrent event; takes two potential values: $D_i(0)$ and $D_i(1)$ \\  
        $U_i$ & $(D_i(0), D_i(1))$; principal stratum membership for individual $i$; takes one of four values: $\nt$, $\co$, $\df$, $\at$  \\  
        $X_i$ & baseline covariates \\
        $C_i$ &  censoring time; takes two potential values: $C_i(0)=C_i(0, D_i(0))$ and $C_i(1) = C_i(1, D_i(1))$  \\
        $T_i$ & failure time; takes two potential values:  $T_i(0) = T_i(0, D_i(0))$  and $T_i(1) = T_i(1, D_i(1))$\\
        $S_z(t)$ & $\Pr(T_i(z) > t)$: potential survival distribution under assignment $z$ \\
        $S_{z,u}(t)$ & $\Pr(T_i(z) > t\mid U_i=u)$: potential survival distribution under assignment $z$ for stratum $u$ in the mixture model approach \\ 
         $S_{z,d}(t)$ & $\Pr(T_i(z) > t\mid D_i=d)$: potential survival distribution under assignment $z$ with  ICE $d$ in the weighting approach \\
          $e(X)$ & $\Pr(Z=1\mid X)$: propensity score  \\
           $p_z(X)$ & $\Pr(D=1\mid Z=z, X)$: conditional probability of the ICE occurrence by treatment arm $z$  \\
           $\pi_u(X)$ & $\Pr(U=u\mid X)$: principal score\\
        \hline
    \end{tabular}
    }
\end{table}

Individual principal stratum membership $U_i$ is not directly observable because it involves the counterfactual ICE status under both assignments, only one of which is observed. Specifically, without additional assumptions, each observed cell of $(Z,D)$ is a mixture of two strata, as shown in the left panel of Table \ref{table:strata-share}. For example, units who were assigned to the treatment arm and continued the treatment ($Z=1,D=1$) can be in the $\at$ or $\co$ stratum. Therefore, the following assumptions are usually made to identify the principal strata estimands.

\begin{ass}[Unconfounded assignment]\label{as:unconfound}
$Z_i \perp \{T_i(0), T_i(1), D_i(0), D_i(1) \} \mid X_i.$
\end{ass}
Assumption \ref{as:unconfound} assumes that the potential outcomes are independent of the assignment conditional on the observed baseline covariates.  This holds by design in randomized trials without conditioning on covariates. Under this setting, the ITT effect is identified by comparing the observed survival function in each arm:
\begin{equation}\label{eq:ITT}
\begin{aligned}
    \tau^{\SPCE}(t) &=\bE[S_1(t)] - \bE[S_0(t)] = \bE[S(t)\mid Z=1] - \bE[S(t)\mid Z=0],
    \end{aligned}
\end{equation}
where $S(t) = \Pr(T > t)$ denotes the \emph{observed} survival distribution of the outcome.

\begin{ass}[Monotonicity]\label{as:monotonicity} $D_i(1)\geq D_i(0)$ for all $i$.
\end{ass}
Monotonicity rules out the $\df$ stratum, as shown in Table \ref{table:strata-share}. Assumption \ref{as:monotonicity} is plausible when the active treatment is more likely to induce ICEs, e.g., when the active treatment has higher toxicity than the control treatment.  Depending on the context and the variable coding, monotonicity can also be $D_i(1)\leq D_i(0)$, that is, when the control arm is more likely to induce ICEs. For example, in some oncology trials, the control arm may be more likely to take subsequent anti-cancer therapy--an ICE--as compared with that in the active treatment arm. Assumption \ref{as:monotonicity} is deterministic, imposed on the individual level; a weaker version is the \emph{stochastic monotonicity} (Assumption 2A), imposed on the distributional level: \cite{cheng2006bounds} $\Pr(D_i(1)=1)\geq\Pr(D_i(0)=1)$.  In a clinical context where the active treatment has higher toxicity than the control treatment, this assumption states that there is a higher probability that patients discontinue the active treatment than the control treatment.

\begin{table}[h]
    \centering
    \setlength\tabcolsep{6pt} 
    \caption{Composition of principal strata in observed cells of assigned and actual treatment $(Z,D)$} \label{table:strata-share}
    \begin{tabular}{c cc|cc}
        & \multicolumn{2}{c}{Without monotonicity} &  \multicolumn{2}{
            |c}{With monotonicity} \\
        \hline
        $Z$ & $D=0$ &$D=1$ & $D=0$ &$D=1$ \\  
        \hline  
        0 & $\nt, \co $ & $\at, \df$  &$\nt, \co $  & $\at$\\
        1 & $\nt, \df$   & $\at, \co$ & $\nt$ & $\at, \co$ \\
        \hline
    \end{tabular}
      \\
      \vspace{3pt}
    \footnotesize{\textbf{Note:}  $\nt$: $U_i=(0,0)$; $\co$: $U_i=(0,1)$; $\at$: $U_i=(1,1) $; $\df$: $U_i=(1,0)$.}
\end{table}

\begin{ass}[Conditionally independent arm-specific censoring]\label{as:censoring}
$T_i(z) \perp C_i(z) \mid 
\{Z_i, U_i, X_i \}$
\end{ass} 
Assumption \ref{as:censoring} states that, conditional on pre-treatment covariates and principal strata, the potential censoring time provides no information on the potential failure time within each arm. 

Note that identification and estimation are two related but distinct concepts. Specifically, identification means that an estimand can be expressed as functions of observed distributions, whereas estimation means using functions of the sample units (i.e. data) to give an informed guess of an estimand. Identification results can help construct estimators and study their properties. Often non-identifiable estimands can still be estimated from data with additional assumptions. Returning to the context of principal stratification, given Assumptions \ref{as:unconfound}-\ref{as:censoring}, there are two strategies to nonparametrically identify the SPCE based on different assumptions, leading to two estimation methods, as well as associated sensitivity analysis. To convey the main message, we will focus on estimation and omit the discussion of identification in the following. Section \ref{sec:ER} and \ref{sec:PI} review these two approaches. Table \ref{table:notations} lists the notations used in the paper.

\subsection{Extension to general intercurrent events} \label{sec:ps-general}
In theory, the principal stratification framework is applicable to any post-treatment variable, but the interpretation of the strata and the choice of the target population depend on the specific clinical context. For example, when the endpoint is progression-free survival time, $D$ indicates the status of cancer progression such as an increase in tumor size, with $D=1$ indicating disease progression and $D=0$ no disease progression. The interpretation of the four strata based on such a $D$ differs from the setting of treatment discontinuation in Section \ref{sec:basic-setup}. In this case, the target estimand would be the difference in the survival time between treatment and control in the stratum of patients whose cancer would not progress regardless of the treatment assignment, i.e. the $U=(D(0)=0, D(1)=0)=00$ stratum. This formulation is the same for the endpoint of complication-free survival time. 

Sometimes, the target population consists of multiple strata. For example, in Bornkamp et.al, \cite{bornkamp2021} the treatment of immunotherapies or targeted anticancer agents may trigger the formation of antidrug antibodies (ADAs), which is an ICE. These ADAs can impact the efficacy and safety of the treatment. A relevant clinical question is whether the ADA-positive patients still benefit from the treatment. Here, the target population is the union of two strata: $\{i: D_i(1) = 1 \} = \{ i: U_i = (0,1)\} \cup \{i: U_i = (1,1)\}$.  

An important choice is the coding of the post-treatment variable. For example, the period that a patient continues to take the assigned treatment in oncology trials is usually measured as a continuous variable, say $\tilde{D}_i$. To simplify the analysis, we often define the discontinuation status $D_i$ by dichotomizing $\tilde{D}_i$ by a cutoff time $d$, e.g., $D_i=1\{\tilde{D}_i<d\}$. The choice of $d$, especially in studies with a long follow-up period as the CALGB 90206 trial, may non-trivially alter the composition of strata. In practice, we recommend conducting a sensitivity analysis on the choice of $d$. 

Principal stratification can be extended to multi-valued ICEs or multiple ICEs. \cite{frumento2012evaluating,mattei2007application} For example, patients may discontinue the assigned treatment or drop out of the study. If we denote the discontinuation and dropout status by $D_1$ and $D_2$, respectively, we can define the principal stratification as $(D_1(0), D_1(1), D_2(0), D_2(1))$, which has $4^2=16$ strata without restrictions. Another example concerns the endpoint of complication-free survival time when both post-treatment complication (denoted by $D_1$) and treatment discontinuation (denoted by $D_2$) are present. A third example is when patients can either switch to the opposite treatment or discontinue both treatments entirely, a common scenario in pragmatic trials. In this case, a three-valued variable $D$ can be defined to encode the ICE, with $D=0, 1, 2$ representing non-switch, switch, discontinuation, respectively, leading to $3^2=9$ strata without restrictions.\cite{liu2024principal}  In general, for a joint principal stratification with $K$ $J$-category ICEs, the total number of strata is $J^{2K}$, increasing exponentially with $K$. Without strong assumptions to reduce the number of strata, each stratum would have too few units to be estimable or interpretable. Therefore, principal stratification is the most suitable for handling a single ICE and it is important to tailor the monotonicity assumption to the specific context based on clinical knowledge to reduce the number of strata.

\section{The Mixture modeling approach} \label{sec:ER}
\subsection{Estimation}\label{subsec:ER-est}
The key challenge to principal strata analysis is that $U_i$ is not observed. To estimate principal causal effects, we need to disentangle the mixtures in the observed cells of $(Z,D)$. Latent mixture modeling naturally lends itself to the task.\cite{imbens1997bayesian, hirano2000assessing} Specifically, we employ two models: (i) the strata model, for predicting the principal strata membership given covariates,  $\pi_u(X) = \Pr(U=u|X)$, also known as the principal score, \cite{jo2009use} and (ii) the outcome model, for predicting the outcomes given the stratum and covariates. Because $U$ is a categorical variable, it is common to posit a multinomial logistic regression model with a reference stratum $u_0$, e.g. the  $\nt$ stratum:
   \begin{equation}\label{eq:umodel}
        \log \bigg\{ \frac{\Pr(U=u \mid X)}{\Pr(U=u_0 \mid X)} \bigg\} = \rho_u+ X' \beta_u,
    \end{equation}
which implies that the principal scores are:
{\small
\begin{align} \label{eq:ps-prop}
    \pi_{u_0}(X)=\Pr(U=u_0 \mid X) &= \frac{1}{1 + \sum_{l\neq u_0} \exp(\rho_l + X' \beta_l)}; \quad  \pi_{u}(X)=\Pr(U=u\mid X) = \frac{\exp(\rho_u + X'\beta_u)}{1 + \sum_{l\neq u_0} \exp(\rho_l + X' \beta_l)}.
\end{align}
}
This multinomial logistic regression model accommodates either 3 or 4 strata, depending on whether the monotonicity assumption \eqref{as:monotonicity} is imposed. 
For time-to-event outcomes, a common choice of the outcome model is the Cox proportional hazard model, \citep{cox1972regression} where the baseline hazard is often assumed to be a Weibull model instead of left unspecified for computational convenience, particularly for Bayesian estimation. Specifically, for stratum $u$ and arm $z$, the hazard function is:   
 \begin{equation}\label{eq:ymodelER}
       h(t; z, u) =  h_0(t; z, u)\exp(X^{'} \gamma_{z,u}) = \exp(\psi_{z,u})t^{\phi_{z,u}-1}\exp(X^{'} \gamma_{z,u})=
t^{\phi_{z,u}-1}\exp(\psi_{z,u} + X^{'} \gamma_{z,u}), \quad \phi_{z,u} > 0. 
    \end{equation}
The corresponding survival functions of the outcome is
\begin{equation}
    S_{z,u}(t \mid X) = \exp\left\{- \int_{0}^{t}h(s;z,u)ds\right\} = \exp\left\{-\frac{t^{\phi_{z,u}} \exp(\psi_{z,u} + X^{'} \gamma_{z,u})}{\phi_{z,u}}\right\}. 
\end{equation}


Either the EM algorithm \cite{dempster1977maximum, zhang2009likelihood} or the Bayesian method \cite{liu2023pstrata} can be employed to handle the unobserved $U_i$ in fitting model \eqref{eq:umodel} and \eqref{eq:ymodelER}. We recommend the Bayesian method for its flexibility and straightforward uncertainty quantification for causal estimands. For Bayesian estimation, we need to specify prior distributions for the model parameters. For example, we can specify standard weakly informative priors: a flat prior for the intercepts, $p(\rho_u) \propto 1$ and $p(\psi_{z,u})\propto 1$, and the log shape parameter in the Weibull model $p(\log \phi_{z,u})\propto 1$; a normal prior for the coefficients: $p(\beta_u) \sim \mathsf{N}(0, \sigma_\beta)$ and $p(\gamma_{z,u})\sim \mathsf{N}(0, \sigma_{\gamma})$, with large prior variance $\sigma_{\beta}$ and $\sigma_{\gamma}$. 

Denote all parameters by $\theta$, Algorithm \ref{alg:mixture} presents the pseudo-algorithm of the Bayesian mixture model with survival outcomes. The algorithm in a wide range of settings has been implemented in the \textbf{PStrata} R package, \cite{liu2023pstrata} leveraging the \textbf{Stan} programming for Bayesian posterior sampling. \cite{Rstan}  More technical details can be found in Liu et al. \cite{liu2024principal} 
\bigskip
\hrule
\begin{algo}\label{alg:mixture}
\textbf{Bayesian mixture modeling estimation of principal causal effects}
\begin{enumerate}
\item Initialize the unknown parameters $\rho_u, \beta_u, \phi_{z,u}, \psi_{z,u}, \gamma_{z,u}$. 
\item For $k=1,\cdots, K$,
\begin{enumerate}
\item Sample the latent strata $U^{k}$ from its posterior predictive distribution based on models (\ref{eq:umodel}) and (\ref{eq:ymodelER}), conditional on the parameter values from the previous iteration ($\theta^{k-1}$).
\item Sample the entire set of continuous parameters $\theta^k = \{\rho^k_{u}, \beta^k_{u}, \phi^k_{z,u}, \psi^k_{z,u}, \gamma^k_{z,u} \text{ for all } z, u\}$ using either a Markov chain Monte Carlo (MCMC) or Hamiltonian Monte Carlo (HMC) step  given the sampled latent strata $U^k$ and observed data.
\item Calculate the survival probability ${S}_{z,u}(t| X; \theta^k)$ and estimate the weighted survival probability $\hat{S}^k_{z,u}(t) = \frac{\sum_{i=1}^{n}\pi_u(X; \theta^k) {S}_{z,u}(t\mid X; \theta^k) }{\sum_{i=1}^{n}\pi_u(X; \theta^k)}$.
\end{enumerate} 
\item Obtain estimates of the target estimand, e.g. the posterior mean and credible intervals empirically from $\{\hat{S}^k_{z,u}(t): k = 1, \cdots, K\}$.
\end{enumerate}
\end{algo}
\hrule

\bigskip

The mixture modeling approach often invokes the exclusion restriction (ER) assumption, \cite{Angrist1996} which posits that the treatment effect in some strata is zero. For example, in the CALGB 90206 trial, one may assume the treatment effect for the $\at$ stratum, i.e., patients who would  discontinue regardless of the assigned treatment, to be zero.
\begin{ass}[Exclusion Restriction]\label{as:ER}
$S_{0,11}(t)=S_{1,11}(t)$ for all $t$. 
\end{ass} 
In model \eqref{eq:ymodelER}, the ER assumption imposes that $( \phi_{0, u}, \psi_{0,u}, \gamma_{0, u}) = (\phi_{1, u}, \psi_{1,u}, \gamma_{1, u})$, for $u = \at$. Due to the complex and diverse reasons for treatment discontinuation, it is not realistic to impose ER for other strata in the CALGB 90206 trial. Both the monotonicity and ER assumptions eliminate certain strata and thus can reduce the uncertainties in the estimation.  \cite{imbens1997bayesian} In the classic setting of treatment switching, Liu et al. \cite{liu2024principal} showed that the SPCE estimands are  nonparametrically identifiable given the monotonicity and ER for two strata.   



\subsection{Characterizing the strata and sensitivity analysis}


Although the individual principal stratum membership is not observed, we can still obtain the summary statistics of each stratum to characterize these latent subpopulations. Specifically, in the Bayesian mixture model approach, we can compute the weighted average of covariates for a given strata $u$ as: $\bar{X}_u = \sum_{i=1}^{n} \widehat{\pi}_{u}(X_i) X_i /\sum_{i=1}^{n} \widehat{\pi}_{u}(X_i)$, where the principal scores $\widehat{\pi}_{u}(X_i)$, namely the probabilities of each unit belonging to each of the possible strata, are calculated based on the fitted model \eqref{eq:ps-prop}. With each posterior draw $k$,
we can compute $\bar{X}^k_u$ in conjunction with $\widehat{S}^k_{z,u}(t)$, as detailed in Algorithm \ref{alg:mixture}, from which we can obtain the posterior distribution of the mean covariate of each stratum. One example is given in Table \ref{tab:stra-X-mixture}.

The monotonicity and ER assumptions are crucial for identifying and inferring the causal estimands, but they cannot be empirically verified using observed data alone. Nonetheless, we can conduct sensitivity analysis to assess the impact of violation to these assumptions. Within the Bayesian paradigm, we can directly compare the posterior distributions of the causal estimands and strata composition when the specific assumption is maintained versus relaxed. \cite{imbens1997bayesian,Mattei2013} 

\section{The Principal Score Weighting approach} \label{sec:PI}
\subsection{Estimation}
The principal score weighting method was first developed by Jiang et al. \cite{jiang2022multiply} for non-censored outcomes. The central identification assumption is the principal ignorability (PI),\cite{jo2009use} whose form depends on the form of monotonicity. Under the monotonicity in Assumption \ref{as:monotonicity}, the PI is as follows. 
\begin{ass}[Principal Ignorability] \label{ass:PI} For all $t \geq 0$, 
$\Pr(T_i(1) \geq t| U_i =\at, X_i) = \Pr(T_i(1) \geq t| U_i =\co, X_i)$ and \\
$\Pr(T_i(0) \geq t| U_i =\co, X_i) = \Pr(T_i(0) \geq t| U_i =\nt, X_i).$ 
\end{ass} 
If the monotonicity is formulated as $D_i(1) \leq D_i(0)$ instead, the corresponding PI is
 $\Pr(T_i(1) \geq t| U_i =\nt, X_i) = \Pr(T_i(1) \geq t| U_i =\df, X_i)$ and 
 $\Pr(T_i(0) \geq t| U_i =\df, X_i) = \Pr(T_i(0) \geq t| U_i =\at, X_i).$
 
PI assumes that, conditional on covariates, the potential survival function under active treatment is identical between the $\at$ stratum and the $\co$ stratum (i.e., those with $D_i(1)=1$), and the potential survival function under control treatment is identical between the $\nt$ stratum and the $\co$ stratum (i.e., those with $D_i(0)=0$). \citep{jo2009use} PI assumes that all the confounding between the strata membership and the outcome is captured by the observed covariates. Therefore, identifying principal causal effects no longer requires conditioning on the unobserved principal stratum, greatly simplifying the analysis. Specifically, under PI, analysts can directly impose models on the observed data $(Z, D, T)$ and use weighting to adjust for confounding covariates to estimate the causal estimands. This underlies the main implementional advantage of the principal score weighting method.  

Cheng et al. \citep{cheng2024multiply} showed that $S_{z,u}(t)$ is nonparametrically identifiable under Assumption \ref{as:unconfound}, \ref{as:monotonicity}, \ref{as:censoring}, \ref{ass:PI}. They proposed a plug-in weighting estimator for SPCE, $\widehat{S}_{z,u}^{\mr}(t)$, which requires specifying and estimating four models. The first model is on the propensity score, $e(X) = \Pr(Z=1\mid X)$, e.g. a logistic regression of $Z$ on $X$,
\begin{equation}\label{eq:ps}
    \logit\{e(X)\} = X' \alpha,
\end{equation}
where $X$ includes an intercept term. In randomized trials, $e(X)$ is known and fixed, e.g., $e(X) = 0.5$ in the CALGB 90206 trial. However, fitting a working model to estimate $e(X)$ can adjust for the potential chance imbalance in the baseline covariates and improve power. \citep{lin2013agnostic} 

The second model is on the principal score, $\pi_u(X)=\Pr(U_i=u\mid X_i)$, which is unobserved because it involves the latent stratum $U$. Instead of directly modeling $\pi_u(X)$, Jiang et al. \cite{jiang2022multiply} suggested to model the probabilities of the \emph{observed} $D$ by treatment arm, $p_z(X)=\Pr(D=1\mid Z=z, X)$, leveraging the following one-to-one relationship between the unobserved scores $\{\pi_{\at}(X), \pi_{\co}(X), \pi_{\nt}(X)\}$ and the observed scores $\{p_0(X), p_1(X)\}$: 
\begin{equation}\label{eq:map}
    \pi_{\at}(X) = p_0(X), \quad \pi_{\nt}(X)= 1-p_1(X),  \quad \pi_{\co}(X) = p_1(X) - p_0(X).
\end{equation}
This relation shows that the principal score is an analogue of the propensity score that summarizes the covariates with respect to the \emph{observed} post-randomization variable $D$. A common model for $p_z(X)$ is a logistic model:
\begin{equation} \label{eq:pmodel}
   \logit\{\Pr(D=1\mid Z=z, X)\} =  X'\phi_z.
\end{equation}

The third and fourth models describe the time-to-event and time-to-censoring processes, respectively. The time-to-censoring model treats censoring as the ``event'' by flipping the event/censoring indicators in the data, i.e., censored observations are marked as events, and actual events are treated as non-events. A common choice is the Cox proportional hazard model for both processes in the observed $(Z,D)$ cell: 
 \begin{align}
       h_{z,d}(t) & =  h_{z,d0}(t)\exp(X^{'} \gamma_{z,d}); \label{eq:ymodelPI} \\
        h_{z,d}^C(t) & =  h_{z,d0}^C(t)\exp(X^{'} \label{eq:cmodelPI}\eta_{z,d}),
    \end{align}
where $h_{z,d0}(t)$ and $h_{z,d0}^C(t)$ are baseline hazard functions of the events and censoring, respectively, for arm $z$ and ICE status $d$. The corresponding conditional survival functions can be obtained as $S_{z,d}(t| X) = \exp\left\{-\int_0^t h_{z,d0}(s)\exp(X'\gamma_{z,d}) \, ds\right\}$ and $S^C_{z,d}(t| X) = \exp\left\{-\int_0^t h_{z,d0}^C(s)\exp(X'\eta_{z,d}) \, ds\right\}$. 

Cheng et al.'s weighting estimator $\widehat{S}_{z,u}^{\mr}(t)$ is a complex function of the estimates of the above four models; the exact form of the estimator is complex and is relegated to the appendix. The estimator is multiply robust in the sense that it is consistent if some set, but not necessarily all, of the four specified models are correctly specified. Algorithm \ref{alg:weighting-MR} outlines the estimation procedure. 
\bigskip
\hrule
\begin{algo} \label{alg:weighting-MR}
\textbf{Multiply-robust weighting estimation of principal causal effects}
   \begin{enumerate}
    \item Obtain the estimated propensity score $\widehat{e}(X) = e(X; \widehat{\alpha})$, where $\widehat{\alpha}$ is the maximum likelihood estimator of the parameters of the propensity score model \eqref{eq:ps}.
    
    \item Obtain estimates of the probabilities $\widehat{p}_z (X)=p_z(X;\widehat{\phi}_z)$, where $\widehat{\phi}_z$ is the maximum likelihood estimator of the parameters of model \eqref{eq:pmodel} based on the observed arm with $Z=z$. Then, using  model \eqref{eq:map}, estimate the principal scores $\widehat{\pi}_{u}(X) = \pi_{u}(X; \widehat{\phi})$ by $\widehat{\pi}_{\at}(X) = p_0(X; \widehat{\phi}_0)$, $\widehat{\pi}_{\nt}(X) = 1- p_1(X; \widehat{\phi}_1)$, and $\widehat{\pi}_{\co}(X) = p_1(X; \widehat{\phi}_1) - p_0(X; \widehat{\phi}_0)$. 
    
    \item  Obtain a doubly robust estimate for $p_z =\bE[p_z(X)]$ as:
    $\widehat{p}_0 = \mP_n[\frac{(1-Z)(D - \widehat{p}_0(X))}{1-\widehat{e}(X)} + \widehat{p}_0(X)]$ and $\widehat{p}_1 = \mP_n[\frac{Z(D - \widehat{p}_1(X))}{\widehat{e}(X)} + \widehat{p}_1(X)]$, where $\mP_n[A]=\sum_{i=1}^{n} A_i/n$ denotes the empirical mean operator. Calculate $\widehat{\pi}_{11}  = \widehat{p}_1, \widehat{\pi}_{00} = 1 - \widehat{p}_1$, and $\widehat{\pi}_{10} = \widehat{p}_1 - \widehat{p}_0$. 
       
    \item Estimate the time-to-event outcome function,  
    $\widehat{S}_{z,d}(t\mid X) = {S}_{z,d}(t\mid X;\widehat{\gamma}_{z,d})$, where $\widehat{\gamma}_{z,d}$ is maximum partial likelihood estimator of $\gamma_{z,d} $ in model \eqref{eq:ymodelPI} based on the observed subjects with $Z = z$ and $D = d$, and the cumulative baseline hazard is estimated via the Breslow estimator.\citep{breslow1972}
    
   \item Analogous to the time-to-event model, obtain estimates of the censoring model \eqref{eq:cmodelPI}, $\widehat{S}^C_{z,d}(t\mid X)= S^C_{z,d}(t\mid X; \widehat{\eta}_{z,d})$.
   
   \item Obtain the plug-in weighting estimates of the survival function $\widehat{S}_{z,u}^{\mr}(t)$ in appendix and consequently the SPCE estimand $\widehat{\tau}^{\SPCE, \mr}_{u}(t)$. 
   
   \item Calculate the standard errors of the estimators via boostrap. 
\end{enumerate} 
\end{algo}
\hrule

\bigskip

\subsection{Characterizing the strata and sensitivity analysis}
Similarly to the mixture modeling approach, strata-specific covariate means can be computed as: $\bar{X}_u = \sum_{i=1}^{n} \widehat{\pi}_{u}(X_i) X_i /\sum_{i=1}^{n} \widehat{\pi}_{u}(X_i) = \mP_n [\widehat{\pi}_{u}(X) X] /\widehat{\pi}_{u} $. The variance is $\Var({X}_u) = \mP_n[(X_i-\bar{X}_u)^2 \widehat{\pi}_{u}(X_i)/\widehat{\pi}_{u}]$.
The principal score model characterizes the ICE-outcome confounding, while the propensity score model in characterizes the covariate-outcome confounding. To evaluate the principal score model, we can assess the balance in covariates before and after the principal score weighting by examining the weighted standardized mean differences (SMDs) of the covariates between the four observed  $(Z, D)$ cell; details are relegated to Appendix. \citep{ding2017principal,cheng2024multiply} In theory, the weighted SMDs converge to zero if the principal score model is correctly specified. Therefore, small weighted SMDs indicate acceptable covariate balance and suggest good fit of the principal score model. 
 
The PI assumption is  central to the weighting method, but it is not directly testable from the observed data. Cheng et al. \cite{cheng2024multiply} designed a sensitivity analysis method for PI with survival outcomes. Specifically, define the following sensitivity functions: 
\begin{equation} \label{eq:PIsensitivity}
\begin{aligned}
    \epsilon_1(t, X) &= \frac{\Pr(T(1) \geq t\mid U = \co, X)}{\Pr(T(1) \geq t\mid U = \at, X)} = \exp\left(\xi_1 \times \frac{t}{t_{\max}}\right), \\
    \epsilon_0(t, X) &= \frac{\Pr(T(0) \geq t\mid U = \co, X)}{\Pr(T(0) \geq t\mid U = \nt, X)} = \exp\left(\xi_0 \times \frac{t}{t_{\max}}\right)
\end{aligned}
\end{equation}
where the two scalar sensitivity parameters $\xi_1, \xi_0$ capture the degree of departure from the PI. When $\xi_1=\xi_0=0$, PI holds; increasing value of $\xi$ implies larger departure from PI. In practice, we can assess the sensitivity to the PI assumption by comparing the weighting estimates $\widehat{\tau}^{\SPCE, \mr}_{u}$ over a plausible range of $( \xi_1, \xi_0)$, which are specified based on domain expertise.

We can also examine the monotonicity assumption \ref{as:monotonicity}. Specifically, define a sensitivity function $\zeta(X) = \pi_{\df
}(X) / \pi_{\co}(X)$ \cite{ding2017principal} to capture the deviation from monotonicity. The ratio lies in $[0,\infty)$, where  $\zeta(X)=0$ corresponds to the monotonicity, and $\zeta(X)=1$ indicates 
 equal proportions of the $\df$ and  $\co$ strata. With fixed $\zeta(X) \neq 1$, the principal scores can be identified as 
\begin{equation} \label{eq:monoweight}
\begin{split}
\pi_{\zeta, \co}(X) & = \{p_1(X)-p_0(X)\}/\{1-\zeta(X)\},  \quad
\pi_{\zeta, \nt}(X)  = 1- p_0(X) - \pi_{\zeta, \co}(X), \\
\pi_{\zeta, \at}(X) &= p_1(X) - \pi_{\zeta, \co}(X), \quad
\pi_{\zeta, \df}(X) = [\zeta(x)\{p_1(X)-p_0(X)\}]/ \{1-\zeta(X)\}.
\end{split}
\end{equation}
For simplicity, we assume that $\zeta(X)$ does not depend on the baseline covariates $X$ and lies between $0$ and $1-\frac{\widehat{p}_1 - \widehat{p}_0}{\min(\widehat{p}_1,1-\widehat{p}_0)}$ to ensure that the estimated proportion of each principal stratum is non-negative.
The sensitivity analysis of monotonicity is applicable to both non-censored and survival outcomes, and can be applied in combination with the sensitivity analysis to the PI assumption.   

\section{Software packages}\label{sec:software}
There are two R packages for principal stratification analysis: 
\begin{itemize}
    \item \textbf{PStrata},\cite{liu2023pstrata} available on CRAN, implements the mixture modeling approach with Bayesian estimation. \textbf{PStrata} is flexible, accommodating outcomes of all types, time-to-event outcomes, different assumptions (e.g., with or without exclusion restriction or monotonicity), multiple binary or categorical ICEs. \textbf{PStrata} accommodates all ICEs that can be formulated as a non-compliance or truncation-by-death problem.    
    \item \textbf{mrPStrata},\cite{cheng2024multiply} available on Github, implements the multiply-robust weighting approach. The current version only accommodates a single binary ICE and requires the monotonicity assumption.    
\end{itemize}
The example R code on synthetic data accompanying this paper is provided in the appendix.

\section{CASE STUDY of the CALGB 90206 Trial}\label{sec:application}

\subsection{Attributes of the estimand}
In line with the ICH E9(R1) addendum, we specify the five attributes of the estimand with a principal strata strategy for ICE in the CALGB 90206 trial. 
\begin{itemize}
    \item[I.] \textbf{Treatment}: (i) Intervention: Bevacizumab (10 mg/kg intravenously every 2 weeks) plus interferon alfa (IFN; 9 million units [MU] subcutaneously three times weekly); (ii) Control: IFN monotherapy (same dose and schedule).  
    \item[II.] \textbf{Population:} the always-continuer stratum of patients, namely, patients aged 18 years and older with metastatic RCC, a clear-cell histologic component confirmed by local pathology review, and no prior systemic therapy for RCC, who would not experienced ICEs regardless of the treatment assignment. 
    \item[III.] \textbf{Endpoint:} the overall survival time, defined as the time from registration to death from any cause.
    \item[IV.] \textbf{Intercurrent events:} treatment discontinuition due to any of the following reasons associated with the clinical questions: 
    (i) never started treatment; (ii) discontinued intervention due to disease progression or toxicity; (iii) refused further treatment. 
    \item[V.] \textbf{Population-level summary:}
     Contrast of the survival probability at each time point between treatment arms in the principal stratum of always-continuer. 
    \end{itemize}
There were six patients whom were lost during the follow-up; our analysis excluded these patients given the small number. Formally, loss-to-follow-up is also an ICE because it prevents measuring the actual treatment status after the initial assignment and leads to censoring of the outcome. It is different from the ICE of treatment discontinuation and requires a separate strategy to handle when its occurrence is suspected to differ between treatment arms. This raises the important question of how to handle multiple ICEs. For the implementional reasons discussed in Section \ref{sec:ps-general}, in this case study, we have pooled discontinuation from different causes into a single binary ICE for illustration, but acknowledge that such a pooling is usually not clinically meaningful.  

\subsection{Data}
The baseline patient information contains demographics (sex and age) and clinical characteristics, including eastern cooperative oncology group performance status, history of nephrectomy and radiation therapy, common sites of metastases, and number of adverse risk factors. The outcome, treatment and baseline covariates are all fully observed. The time off the assigned treatment is recorded for each patient, regardless of the reason of discontinuation. The $25\%, 50\%, 75\%$, and $100\%$ quantiles of the discontinuation time, excluding 49 missing cases, are 60.5, 111, 277.5, and 1,454 days, respectively, with a mean of 111 days. Because nearly all participants discontinued treatment at some point before the designated end of the study, and given the delayed therapeutic effects of the drugs, we define the ICE by dichotomizing the discontinuation time using a prespecified cutoff point $d$: the occurrence of ICE (discontinuation, $D=1$) is defined as if the time off treatment attributable to any of the four predefined reasons is shorter than $d$ days, and no ICE (continuation, $D=0$) otherwise.  Table \ref{tab:comp-real} in Appendix C provides the composition of $(Z, D)$ under different cutoff values of $d$.
For illustration, we chose the discontinuation cutoff at $d= 90$.  Because of the high toxicity and rapid disease progression observed in the trial, we assume that the discontinuation occurred within 90 days ($D=1$) for participants whose time off assigned treatment is missing. 

Table \ref{tab:baseline} summarizes the baseline characteristics by randomization arm and discontinuation status. Most covariates are well
balanced between the arms, but there are notable differences in some covariates between the continued and discontinued subgroups. For example, the prevalence of nephrectomy and having more than two risk factors is significantly higher in the discontinued subgroup than in the continued subgroup, regardless of the treatment arm. This indicates that discontinuation is likely confounded.

\subsection{Results and sensitivity analysis of the mixture modeling approach}
This section re-analyzes the CALGB 90206 trial with the Bayesian mixture model approach, using the \textbf{PStrata} \citep{liu2023pstrata} package; the R code is given in Appendix A.  Assumption \ref{as:unconfound} holds by the randomization design and Assumption \ref{as:censoring} is deemed reasonable. However, the ER and monotonicity assumptions are not plausible in this trial. We fit the multinomial strata model \eqref{eq:umodel} and the Weibull-Cox outcome model \eqref{eq:ymodelER} to the data. We incorporate covariates that are clinically considered predictive of the discontinuation status or outcome, including age, sex, history of nephrectomy (1 = yes), and status, defined as having more than two adverse risk factors (1= yes).
We run three parallel Markov chains, each with 10,000 iterations including 5,000 burn-in iterations. The estimated stratum composition is reported in Table~\ref{tab:strata-mixture}, and the stratum-specific distribution of covariates is shown in Table~\ref{tab:stra-X-mixture}. These results reveal substantial differences between the principal strata, suggesting a meaningful separation. Specifically, the \emph{always-continuer} ($\nt$) and \emph{always-discontinuer} ($\at$) strata contain a higher proportion of male participants. Also, the always-discontinuer stratum has the highest prevalence of nephrectomy and $\geq 3$ adverse risk factors, whereas the $\df$ stratum has the lowest prevalence of $\geq 3$ adverse risk factors.

\begin{figure}[h]
  \centering
    \includegraphics[width=0.9\linewidth]{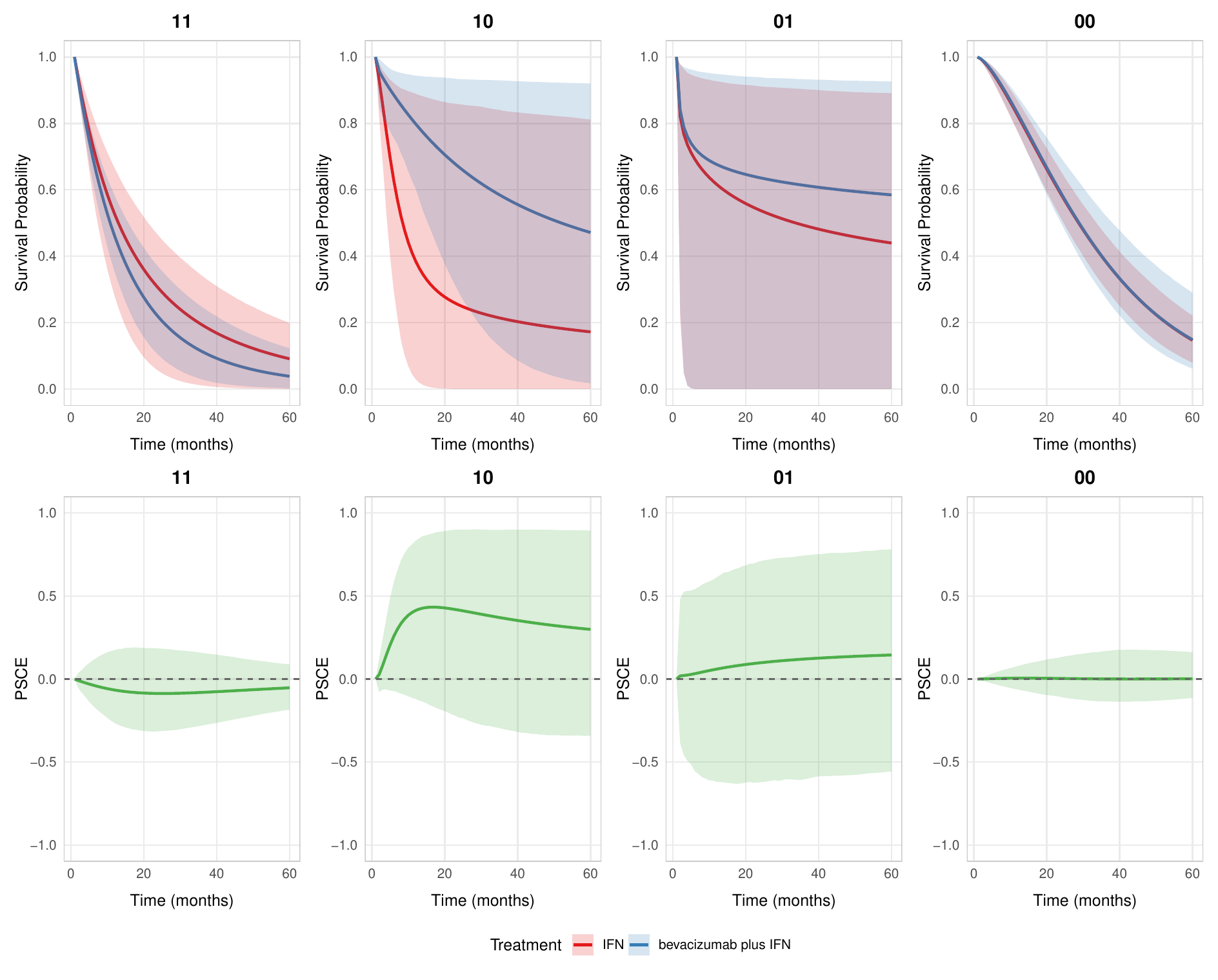} 
    \caption{Posterior survival probability curves and treatment effects of each principal stratum in the CALGB 90206 trial, obtained via the Bayesian mixture model method without the monotonicity or ER assumption. }
   \label{fig:SPCE-Mix}
\end{figure}

The survival probabilities and SPCEs estimated via the mixture model approach with either the monotonicity or the ER assumption are presented in Figure \ref{fig:SPCE-Mix}. The results suggest heterogeneity in treatment effects between principal strata. Specifically, assignment to bevacizumab plus IFN shows a negligible effect throughout the follow-up period for the target subpopulation of \emph{always-continuer} (the $\nt$ stratum), a protective  effect for the $\co$ and $\df$ strata, and a harmful effect for the always-discontinuer ($\at$) stratum. For the $\co$ stratum, the protective effect increases over time, peaking at 50 months with a SPCE of $0.12$ with 95\% interval (-0.65, 0.77).  For the $\df$ and $\at$ strata, the magnitude of the SPCE increases initially and then decreases over time, peaking at 14 months with an SPCE of $0.39 (-0.21, 0.90)$ for the $\df$ stratum, and at 22 months with an SPCE of $-0.06 (-0.30, 0.18)$ for the $\at$ stratum. However, all estimates are associated with large uncertainties, with wide credible intervals that include zero. This is not surprising given the lack of restricting assumptions (see Imbens and Rubin \cite{imbens1997bayesian} for more examples and discussion).

Notably, the proportion of the $\co$ stratum is small (4$\%$). This motivates us to conduct a sensitivity analysis on the monotonicity assumption to rule out the $\co$ stratum. The results are delegated to Figure \ref{fig:SPCE-Mix-Mono} in Appendix C.  In this setting, 
the proportion of the $\df$ stratum decreases from $12.4\% $  to $7.9\%$, while the proportions of the $\nt$ and $\at$ strata increase. The point estimates of the SPCEs in the three strata are similar to those estimated without the monotonicity assumption. As expected, monotonicity leads to tighter credible intervals, reducing the SPCE intervals by 50\%, but all intervals still contain zero. 

To further illustrate the impact of the ER assumption, we conduct two additional sensitivity analyses: impose the ER assumption to the \emph{always-discontinued} ($\at$) stratum, without and with monotonicity, the results of which are given in Figure \ref{fig:SPCE-Mix-ER} and \ref{fig:SPCE-Mix-Mono-ER} in Appendix C, respectively. The proportions of the strata and the stratum-specific summary of covariates under all four combinations of monotonicity and ER are given in Table~\ref{tab:strata-mixture} and \ref{tab:stra-X-mixture-SA} in Appendix C, respectively. We can see that, without monotonicity, imposing the ER assumption has little impact on the stratum proportions and the SPCE estimates. 
With monotonicity, the proportion of $\df$ stratum slightly decreases while the proportion of  both the $\nt$ and $\at$ strata increases. Importantly, point estimate of the SPCEs and the stratum-specific summary of covariates show little variation between the four scenarios, supporting the robustness of the findings.

\subsection{Results and sensitivity analysis of the principal score weighting approach}
\begin{figure}
    \centering
    \includegraphics[width=0.9\linewidth]{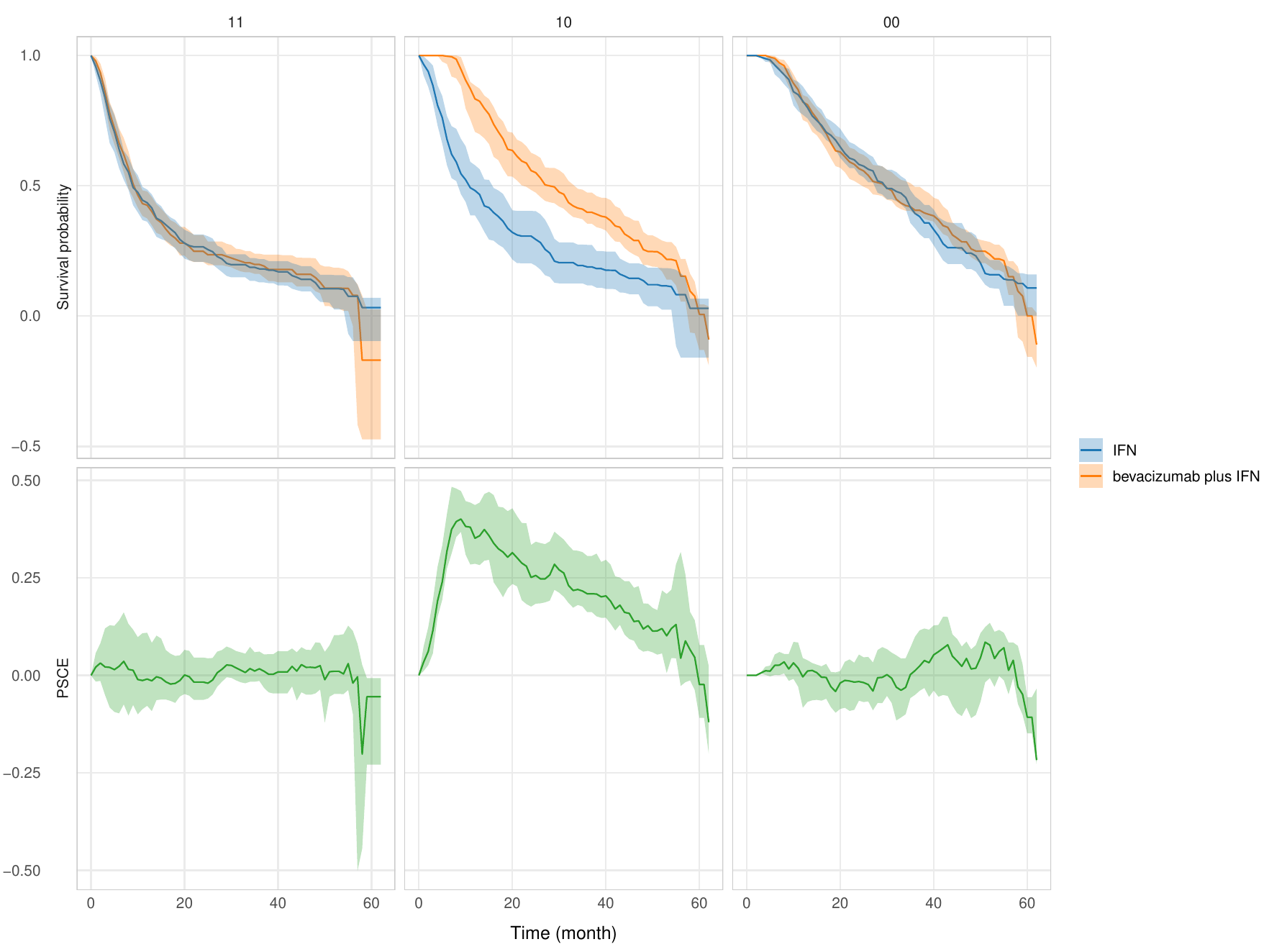} 
    \caption{Estimated survival probability curves and treatment effect  using the weighting method in the CALGB 90206 trial. } \label{fig:SPCE-PI}
\end{figure}

This section re-analyzes the CALGB 90206 trial with the multiply robust weighting estimator, using the \textbf{mrPStrata} package; the R code is given in Appendix A. 
We first assumed the monotonicity and PI assumptions as the benchmark, resulting in three principal strata  $\nt$, $\df$, and $\at$, with
estimated proportions of 0.45, 0.13, and 0.42, respectively. In Appendix C, we provide the stratum-specific covariate means in Table \ref{tab:stra-X-weighting}, which are similar to those obtained in the mixture model method. We examine the SMD of the covariates between arms (Figure \ref{fig:SMD} in Appendix C) and find all unweighted and weighted SMDs are below 0.2, indicating satisfactory covariate balance between the comparison groups: $(Z=1, D=1)$ vs. $(Z=0, D=0)$, $(Z=1, D=0)$ vs. $(Z=0, D=0)$, and $(Z=1, D=1)$ vs. $(Z=0, D=1)$. Figure \ref{fig:SPCE-PI} presents the SPCE estimates and the counterfactual survival functions of each principal stratum.  The point-wise $95\%$ confidence intervals are obtained using the nonparametric bootstrap with $1,000$ replicates. 
The trend in the survival curves is consistent with that from the mixture model approach, but the confidence bands are much narrower because of the PI assumption. 
For the \emph{always-continuer} ($\nt$) stratum, assignment to the bevacizumab plus IFN arm results in higher survival probabilities at most time points; however, the corresponding confidence intervals still include zero throughout the follow-up period. For the $\df$ stratum, bevacizumab plus IFN demonstrates a consistently significant and strong protective effect on survival across time, peaking at approximately 10 months with an estimate of $0.47 (0.18, 0.58)$. For the \emph{always-discontinuer} ($\at$) stratum, the effects fluctuate over time, oscillating between relatively small positive and negative values within the range of $(-0.05, 0.05)$.

\begin{figure}
    \centering
    \includegraphics[width=0.9\linewidth]{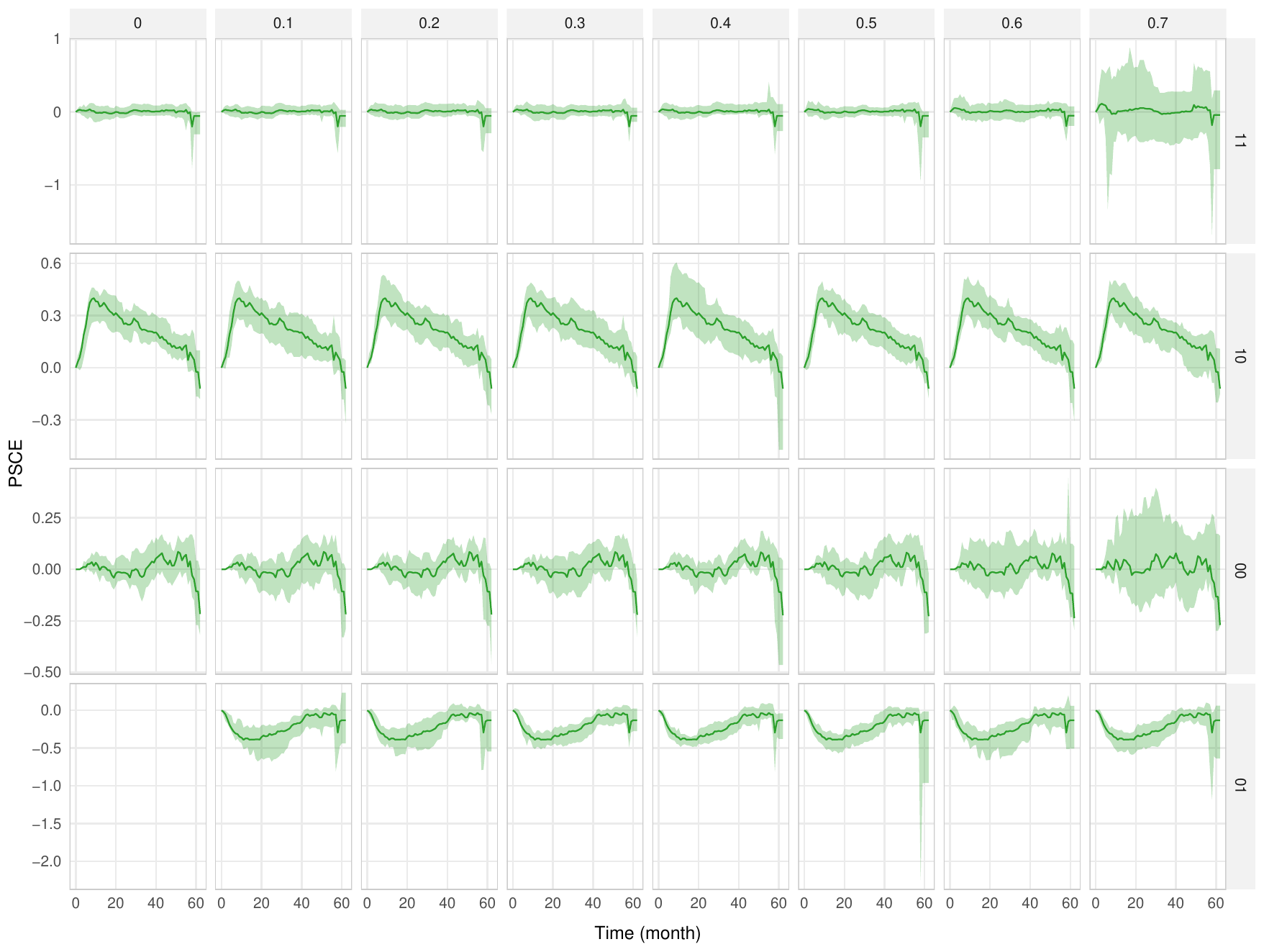} 
    \caption{The PSCE estimates in the sensitivity analysis of the monotonicity assumption with different $\zeta(X)$ values under the weighting method in the CALGB 90206 trial. } \label{fig:SPCE-PI-mono}
\end{figure}

To investigate the sensitivity to the monotonicity assumption in the weighting approach, we explore  $\zeta(X) = \{0, 0.1, 0.2, 0.3, 0.4, 0.5, 0.6, 0.7\} $
over the range $[0, 1-\frac{\widehat{p}_1 - \widehat{p}_0}{\min(\widehat{p}_1,1-\widehat{p}_0)}] = [0, 0.76]$ and present the  results in Figure \ref{fig:SPCE-PI-mono}. The results suggest that the SPCE estimates for the  $\at$, $\df$, and $\nt$ strata are robust when the $\co$ stratum exists. The SPCE in the 
$\co$ stratum is consistently significantly negative before 60 months, with the largest effect of approximately $-0.4$ occurring around month 10, regardless of the value of $\zeta(X)$. This suggests that the intervention has a harmful effect on the survival of patients who would discontinue treatment in the intervention arm but would continue treatment in the control arm. Furthermore, to investigate the PI assumption, we vary the sensitivity parameters $\{\xi_0, \xi_1 \}$ in Equation \ref{eq:PIsensitivity} over the grid  $\{\log 0.9, \log 1.2, \log1.5 \} \times \{\log 0.9, \log 1.2,  \log1.5  \}$. The SPCE estimates, shown in Figure~\ref{fig:SPCE-PI-sen} in Appendix C, remain stable across the range, with only slight variation in the confidence intervals.

In summary, we applied both the Bayesian mixture model and the weighting methods within the principal strata strategy to address the ICE of treatment discontinuation in the CALGB 90206 trial. We conducted extensive sensitivity analysis to examine potential violations of the key assumptions. Our results suggest: (i) there is heterogeneity in the treatment effects of the intervention  between different principal strata; (ii) for the stratum of patients who would continue the assigned treatment regardless of the initial assignment, there is a null effect of the intervention, (iii) the qualitative conclusions are consistent between the two estimation methods, but the weighting method leads to much higher precision than the mixture model method, and (iv) the results are robust to violations of the key causal assumptions in each method, including monotonicity, PI and ER.

\begin{table}[ht]
\centering
\caption{Baseline characteristics by randomized arm and discontinuation status in the CALGB 90206 Trial}
\resizebox{\textwidth}{!}{%
\begin{tabular}{@{}lccccccccccccccc@{}}
\toprule
& \multicolumn{3}{c}{\textbf{IFN} ($Z=0$)} & \multicolumn{3}{c}{\textbf{Bevacizumab plus IFN} ($Z=1$)} \\
\cmidrule(lr){2-4}  \cmidrule(lr){5-7}
\textbf{Characteristic} & \textbf{Overall} & \shortstack{\textbf{Continued} \\ ($D=0$)} & \shortstack{\textbf{Discontinued} \\ ($D=1$)} & \textbf{Overall} & \shortstack{\textbf{Continued} \\ ($D=0$)} & \shortstack{\textbf{Discontinued} \\ ($D=1$)} \\
\midrule
Male & 239 (65.8$\%$) & 108 (65.9$\%$) & 131 (65.9$\%$)  & 269 (63.4$\%$)  & 154 (72.0$\%$) & 115 (74.2$\%$) \\
Nephrectomy & 55 (15.2$\%)$ & 17 (10.4$\%$)  & 38 (19.1$\%$) &57 (15.4$\%)$  & 29 (13.6$\%$)  & 28 (18.1$\%$)  \\
0 risk factors & 95 (26.2$\%$) & 49 (29.8$\%$)  & 46 (23.1$\%$) & 97(26.3$\%$) & 65 (30.4$\%$) & 32 ((20.7$\%$)\\
1-2 risk factors  & 231 (63.6$\%$) & 104 (63.4$\%$) & 127 (63.8$\%$) &234 (63.4$\%$)  & 137 (64.2$\%$) & 97(62.6$\%$) \\
$\geq 3$ risk factors & 37 (10.2$\%$) &  11 (6.7$\%$) & 26 (13.1$\%$)  & 38 (10.3$\%$) & 12 (5.7$\%$)  & 26 (16.8$\%$)  \\
Age & 61.9 (21, 83)  & 61.2 (30, 81) & 62.4 (21, 83)  & 62.1 (36, 89)  &61.5 (37, 88)  & 63.0 (36, 89)  \\
\bottomrule
\end{tabular}\label{tab:baseline}
}
\end{table}

\begin{table}[]
    \centering 
        \caption{Estimated proportions of strata in the CALGB 90206 trial under different combinations of the monotonicity and ER assumptions}
    \label{tab:strata-mixture}
    \resizebox{\textwidth}{!}{%
    \begin{tabular}{cccccccccccccc}
    \hline
 \toprule
\textbf{Assumption} & \multicolumn{2}{c}
{$\mathbf{U=(0,0)}$} & \multicolumn{2}{c}{$\mathbf{U=(1,0)}$} & \multicolumn{2}{c}{$\mathbf{U=(1,1)}$} & \multicolumn{2}{c}{$\mathbf{U=(0, 1)}$}\\
 & \textbf{Mean} & \textbf{95\% CI} & \textbf{Mean} & \textbf{95\% CI} & \textbf{Mean} & \textbf{95\% CI} & \textbf{Mean} & \textbf{95\% CI}\\
 \midrule
\textbf{Monotonicity, no ER ($\%$) } & 48.1 & (42.9, 53.0)  &  7.9  & (1.0, 15.5) & 44.0 &(38.9, 49.3)    &-- &-- \\ 
\textbf{Monotonicity, ER ($\%$)} & 45.1 & (40.1, 49.9) & 5.7 & (0.6, 12.6) & 49.2 & (44.1, 53.9)  &-- &--\\ 
\textbf{No monotonicity, no ER ($\%$)} & 44.0 & (35.9, 50.3) & 12.4 & (3.7, 23.3) & 39.6 & (31.7, 47.0) 
 & 4.0 & (0.4, 8.7) \\ 
 \textbf{No monotonicity, ER ($\%$)} & 44.2 & (35.8, 50.6) & 11.8 & (2.8, 23.0) & 39.8 & (30.1, 47.6) 
 & 4.2 & (0.4, 9.7)\\
 \hline         
    \end{tabular}
    }
\end{table}

\begin{table}[ht]
\centering
\caption{Stratum-specific summary of covariates in the CALGB 90206 trial using the Bayesian mixture model without assuming monotonicity or ER}
\resizebox{\textwidth}{!}{%
\begin{tabular}{lcccccccc}
\toprule 
\textbf{Variable} 
& \multicolumn{2}{c}{$\mathbf{U=(0,0)}$} & \multicolumn{2}{c}{$\mathbf{U=(1,0)}$} & \multicolumn{2}{c}{$\mathbf{U=(1,1)}$} & \multicolumn{2}{c}{$\mathbf{U=(0,1)}$}\\
 & \textbf{Mean} & \textbf{95\% CI} & \textbf{Mean} & \textbf{95\% CI} & \textbf{Mean} & \textbf{95\% CI}& \textbf{Mean} & \textbf{95\% CI} \\
\midrule
Male ($\%$) & 70.1 & (65.6, 74.5)  &  66.3 & (44.7, 84.1) & 70.0 & (64.4, 75.3) & 57.2& (18.1, 82.2) \\
Nephrectomy ($\%$) & 12.3  & (9.0, 15.3) & 11.6 & (2.6, 24.5) & 20.6  & (16.8, 25.0) &  8.2 & (1.4,  23.8) \\
0 risk factors ($\%$) & 27.5 & (26.5, 28.3) & 28.5 & (24.7, 31.5)   & 24.0 & (22.5, 25.1)  &  29.7 & (24.9, 33.7) \\
1-2 risk factors ($\%$) & 66.0 & (64.4, 67.6) & 66.4 & (60.5, 70.2)   & 59.6 & (57.0, 62.0)  &  64.9 &(57.0, 69.1) \\
$\geq 3$ risk factors ($\%$)& 6.5  &(4.3, 8.9)  & 5.1  &(0.7, 14.0)  & 16.4  &(13.1, 20.3)   & 5.4 &(0.9,  17.0)\\
Age (years) & 61.1  &(60.1,  62.1) & 63.1  &(58.9,  67.5) &   62.3  &(61.1,  63.6) & 64.3 & (56.0, 70.7)\\
\bottomrule
\end{tabular}\label{tab:stra-X-mixture}
}
\end{table}

\section{Trial-Based Simulation Analysis}\label{sec:simu}
To further investigate the operational characteristics of the principal stratum strategy in realistic settings, we perform simulation studies based on the CALGB 90206 trial. 
We generate 732 patients and randomly assign $363$ and $369$ of these patients to the treatment ($Z=1$) and control ($Z=0$) arm, respectively. For each patient, we simulate four baseline covariate $X$, mimicking age, gender, history of nephrectomy, and number of adverse risk factors, 
using the summary statistics (means, proportions, and variance) in Table 1 of Rini et al.\cite{rini2008bevacizumab}. These covariates are then recoded to match the specifications used in the case study.

We generate the four principal strata $(U \in \{ \nt, \df, \at, \co \})$ using model \eqref{eq:umodel}
with $\beta_{\df} = (-1.4, 0, 0.1, 0.2, 0.5)$, $\beta_{\at} = (0, -0.1, -0.1, -0.1, 0)$ and $\beta_{\co} = (-2.3, -0.1, -0.1, -0.1, 0.5)$. This results in proportions of $0.44, 0.12, 0.40$ and $0.04$, respectively, matching the estimated proportions of the four principal strata in the case study. We generate the outcome models for each stratum $u$ and treatment arm $z$ using \eqref{eq:ymodelER}, where the coefficients $\phi_{z, u}$, $\psi_{z,u}$ and $\gamma_{z, u}$ are shown in Table \ref{tab:sim_Y} in Appendix D. In this setting, neither the ER nor the PI assumption holds exactly, but the deviation from these assumptions is not severe.
The censoring is drawn independently from an exponential distribution with rate $\exp(-5 - X_1 - X_2 - X_3 - X_4)$, leading to an overall censoring rate of 0.1, similar to the CALGB 90206 trial.


We first fit the mixture model without either the ER or the monotonicity assumption. We run 6 chains, each with 2,000 iterations including 1,000 burn-in iterations. Both the estimated stratum probabilities and the stratum-specific causal effects have wide confidence intervals. Specifically, the estimated stratum probabilities are 46.2\% (0.4\% -- 58.7\%)  for $U = \nt$, 8.8\% (0.5\% -- 49.3\%) for $U = \df$, 36.2\% (0.4\% -- 46.9\%)  for $U = \at$, and 8.8\% (0.1\% -- 52.8\%) for $U = \co$, respectively. These wide intervals reflect the intrinsic difficulty in disentangling the components from a mixture without additional assumptions. The large uncertainty in identifying the strata renders the stratum-specific causal effects not informative.

\begin{figure}[h]
    \centering
    \includegraphics[width=0.9\linewidth]{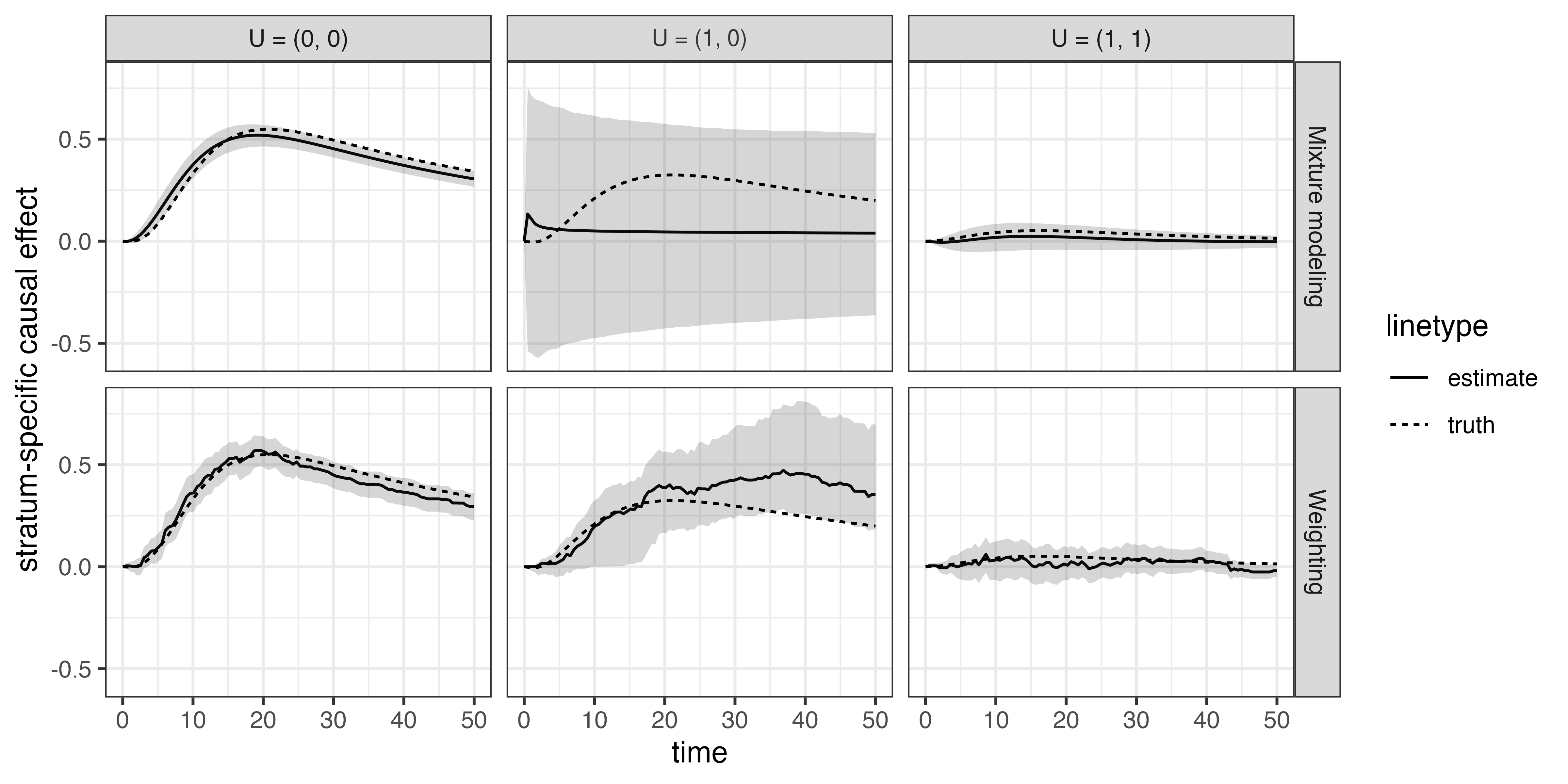}
    \caption{Fitted stratum-specific causal effects in the simulation study with the mixture model and weighting methods, respectively. Both methods assume the monotonicity assumption. The true causal effects are provided in dashed lines for reference.}
    \label{fig:sim-causal-effect}
\end{figure}

We rerun the analysis assuming monotonicity, which rules out the $\co$ stratum. The proportions of the strata are 51.1\% (47.6\% -- 54.7\%) for $U = \nt$ , 0.9\% (0.3\% -- 2.4\%) for $U = \df$, and 48.0\% (44.4\% -- 51.5\%) for $U = \at$, respectively. The stratum-specific causal effects are shown in the upper panel of Figure \ref{fig:sim-causal-effect}. Specifically, the estimated causal effect of the target $\nt$ stratum, is very close to the truth (in dashed lines). Furthermore, the estimated causal effect of the \emph{11} stratum is close to zero, suggesting that the ER assumption (Assumption \ref{as:ER}) likely holds. However, the estimated causal effect of the $\df$ stratum is associated with a wide confidence interval, making the estimates not informative. This is not surprising given the small proportion of units estimated to be in the stratum. We further conduct an analysis assuming both monotonicity and ER; the results are almost identical and are thus omitted.

We then analyze the simulated data with the weighting method, assuming the monotonicity and PI assumptions. We obtain the 95\% confidence interval by bootstrapping 1000 times. The proportions of the principal strata are estimated to be 45.0\% (40.0\% -- 50.1\%) for $U=\nt$,  11.7\% (6.7\% -- 19.2\%) for $U = \df$, and 43.5\% (38.4\% -- 48.6\%) for $U = \at$ respectively.  The estimated and true stratum-specific causal effects are shown in the lower panel of Figure \ref{fig:sim-causal-effect}. For both the $\nt$ and $\at$ strata, the estimated causal effects are close to the truth with relatively narrow confidence bands. The confidence interval for the $\df$ stratum is still wide, but does not cover 0, due to the much larger estimated proportion than that in the mixture model method.

The simulations show that both the mixture model and the weighting methods for principal stratification, despite relying on different assumptions, can recover the true principal causal effects for the strata with enough sample units (e.g. the $\nt$ and $\at$ strata), even when some of the causal assumptions are mildly violated. But for strata with few patients, both methods lead to unreliable results.  

\section{CONCLUSION}\label{sec:conclusion}
We presented a tutorial on the principal stratum strategy to handle intercurrent events in studies with survival outcomes within the Estimands framework for clinical trials. The principal stratum strategy is conceptually and technically complex, hindering its adoption in practice. We reviewed basic concepts, estimation, and programming using an oncology clinical trial with substantial incidence of treatment discontinuation as an illustrative case study. We provided example R codes within existing R packages for implementing the methods. 

We review two estimation methods for the principal stratum strategy: the Bayesian mixture modeling and the frequentist weighting methods. Each method relies on a set of different causal assumptions. The Bayesian method is flexible to incorporate prior knowledge and different assumptions as well as conduct sensitivity analysis, but it requires intensive computational resources and can be sensitive to model specification. In contrast, the weighting method is more robust to model specification and computationally more efficient, but it requires the principal ignorability assumption, which is not always plausible; also, the current software for the weighting method requires the monotonicity assumption.  
The choice of an estimation approach in a specific study should be guided by the clinical context and the plausibility of the underlying assumptions. In practice, we recommend analysts to analyze their study using both methods when possible: similar results between methods would lend more credibility to the conclusions; vastly different results would suggest violation of causal or model assumptions. Because principal strata are latent, it is crucial to characterize the covariates profile of the strata in the analysis. 

The principal strata strategy, when properly implemented, provides additional information on the treatment effect heterogeneity in the presence of ICEs that is not captured by the standard ITT analysis. However, our case study demonstrates a common problem of this strategy: the principal effects are often estimated with large uncertainties even in the simplest case of a single binary ICE, leading to inconclusive results. Imposing structural assumptions such as monotonicity, exclusion restriction, and principal ignorability can reduce the uncertainties. However, these assumptions are not always plausible in real studies. Investigators should carefully choose assumptions based on clinical knowledge and routinely perform sensitivity analysis to the key assumptions, as recommended by the ICH E9 (R1) addendum. Robustness to violations of causal assumptions adds credibility to the results, whereas increased sensitivity to one or more assumptions demands more caution in interpreting the results. 

This tutorial focuses on the type of ICE that can be formulated into the non-compliance setting within principal stratification. Another common type of ICE is \emph{truncation by death},\cite{zhang2003estimation,Rubin06} where the occurrence of an intercurrent event, e.g. death, precludes complete or partial observation of the outcome of interest, e.g. quality of life, and the primary causal estimand is the survivor average causal effect (SACE). In terms of data structure, the main difference between the noncompliance and truncation-by-death settings is that conceptually outcomes are observed in all units in the former but are only observed on a subset of units in the latter.   Therefore, causal inference in these two settings rests on different assumptions and inferential procedures.\cite{ding2011identifiability,tchetgen2014identification} The \textbf{PStrata} package accommodates both settings and the accompanying manual \cite{liu2023pstrata} provides implementation details.

\section*{Acknowledgment}
This research is funded by the Food and Drug Administration (FDA) of the U.S. Department of Health and Human Services (HHS) as part of a financial assistance award Center of Excellence in Regulatory Science \& Innovation, U01FD007857 totaling \$673,975 from Oncology Center of Excellence (OCE). The contents are those of the author(s) and do not necessarily represent the official views of, nor an endorsement, by FDA/HHS, or the U.S. Government.

\bibliographystyle{unsrtnat}
\bibliography{oncology}

\clearpage
\newpage

\clearpage
\newpage

\clearpage
\newpage

\clearpage
\newpage
\section*{Appendix}\label{sec:appendix}
\subsection*{A. Example R code with the CALGB 90206 trial data}
\hrule
\begin{lstlisting}[style=simTutorialR, caption={Code for the mixture modeling approach using R package \textbf{PStrata}}, label={lst:pstrata}]
# Install and load package ---------------------------------------------------
install.packages("PStrata")
library(PStrata)
library(ggplot2)

# Data generation code is available at: 
# https://github.com/xiaoxiaoduke/PS-Survival-code

# Load data----------------------------
data <- readRDS("data.rds")
# Specify model formulas -----------------------------------------------------
S.formula <- arm + D ~ ages + SEX + nephrectomy + riskfac
Y.formula <- survtime + death ~ ages + SEX + nephrectomy + riskfac

# Fit principal stratification model  ------------------------
fit <- PStrata(
  S.formula         = S.formula,
  Y.formula         = Y.formula,
  Y.family          = survival(method = "Cox"),
  data              = data,
  strata            = c(a = "11", n = "00", c = "10"), # Monotonicity, no ER
  # strata            = c(a = "11*", n = "00", c = "10"), #  Monotonicity,  ER 
  # strata            = c(a = "11", n = "00", c = "10", d = "01"), # No monotonicity, no ER
  # strata            = c(a = "11*", n = "00", c = "10", d = "01"), # No monotonicity,  ER 
  prior_intercept   = prior_normal(0, 1),
  prior_coefficient = prior_normal(0, 1),
  warmup            = 2000,
  iter              = 4000,
  chains            = 3
)

# Summarize fitted model -----------------------------------------------------
summary(fit)   # stratum-specific summaries

# Posterior predictive survival probability curves --------------------------------------
fit_outcome <- PSOutcome(fit)
plot(fit_outcome) + 
  xlab("time") + 
  ylab("survival probability")

# Extract outcome array ------------------------------------------------------
outcome_array <- fit_outcome$outcome_array
\end{lstlisting}
\hrule

\newpage
\hrule
\begin{lstlisting}[style=simTutorialR, caption={Code for the weighting approach using R package \textbf{mrPStrata}}, label={lst:pstrata}]
# Install and load package ---------------------------------------------------
devtools::install_github("chaochengstat/mrPStrata")
library(mrPStrata)

# Load data-----------------------------
data <- readRDS("data.rds")

# Implement the multiply robust estimators------------------------------------
res = mrPStrata(times=seq(0,7,0.1),
                data = data,
                Xpi_names = c("ages","SEX","nephrectomy","riskfac"),
                Xe_names = c("ages","SEX","nephrectomy","riskfac"),
                Xc_names = c("ages","SEX","nephrectomy","riskfac"),
                Xt_names = c("ages","SEX","nephrectomy","riskfac"),
                Z_name = "arm",
                S_name = "D", 
                U_name ="survtime",
                delta_name = "death",  
                B=50)

# Extract results by stratum -------------------------------------------------
print(res$Compliers) 
print(res$Always_Takers)
print(res$Never_Takers)

# Visualize results ----------------------------------------------------------
plot.psce(res)

#Sensitivity analysis for the monotonicity------------------------------------
res_mo = mrPStrata_MO_SA(times=seq(0,7,0.1),
                data = data,
                Xpi_names = c("ages","SEX","nephrectomy","riskfac"),
                Xe_names = c("ages","SEX","nephrectomy","riskfac"),
                Xc_names = c("ages","SEX","nephrectomy","riskfac"),
                Xt_names = c("ages","SEX","nephrectomy","riskfac"),
                Z_name = "arm",
                S_name = "D", 
                U_name ="survtime",
                delta_name = "death",  
                zeta=0.2,
                B=50)
                
#Sensitivity analysis for the principal ignorability--------------------------
res_PI = mrPStrata_PI_SA(times=seq(0,7,0.1),
                data = data,
                Xpi_names = c("ages","SEX","nephrectomy","riskfac"),
                Xe_names = c("ages","SEX","nephrectomy","riskfac"),
                Xc_names = c("ages","SEX","nephrectomy","riskfac"),
                Xt_names = c("ages","SEX","nephrectomy","riskfac"),
                Z_name = "arm",
                S_name = "D", 
                U_name ="survtime",
                delta_name = "death",  
                xi0 = log(0.9),
                xi1 = log(0.9),
                eta0=1,
                eta1=1,
                B=50)
\end{lstlisting}
\hrule

\subsection*{B. Plug-in multiply-robust weighting estimator $\widehat{S}_{z,u}^{\mr}(t)$}

\begin{equation}
{\small
    \begin{aligned}
    \widehat{S}^{\mathrm{mr}}_{0,\co}(t) & = \mP_n \Bigg\{
\frac{\widehat{\pi}_{\co}(X)}{\widehat{p}_1 - \widehat{p}_0}
\frac{1-D}{1-\widehat{p}_0(X)} 
\frac{1-Z}{ 1-\widehat{e}(X)}
\bigg[ \frac{\mathbb{I}(C \geq t)}{\widehat{S}^{C}_{0,0}(t|X) } 
+ \widehat{S}_{0,0}(t|X) 
\int_0^t \frac{d\widehat{M}_{0,0}^{\mathrm{C}}(r|X)}{\widehat{S}_{0,0}(r|X) \widehat{S}^{\mathrm{C}}_{0,0}(r|X)} \bigg] 
\\
& + \frac{\widehat{S}_{0,0}(t|X)}{\widehat{p}_1 - \widehat{p}_0} \bigg[\frac{Z}{\widehat{e}(X)}( D - \widehat{p}_0(X))
+ \frac{1-\widehat{p}_0(X)}{1-\widehat{p}_1(X)} \frac{1-Z}{1-\widehat{e}(X)}  (\widehat{p}_1(X) - D)\bigg]
 + \frac{\widehat{\pi}_c(X)}{\widehat{p}_1 - \widehat{p}_0} 
\bigg( 1 - \frac{1-Z}{1-\widehat{e}(X)} \bigg) 
\widehat{S}_{0,0}(t|X)
\Bigg\}, 
\end{aligned}
}
\end{equation}
{\small
\begin{align}
    \widehat{S}^{\mathrm{mr}}_{1,\co}(t) & = \mP_n \Bigg\{
\frac{\widehat{\pi}_{\co}(X)}{\widehat{p}_1 - \widehat{p}_0}
\frac{D}{\widehat{p}_1(X)} 
\frac{Z}{ \widehat{e}(X)}
\bigg[ \frac{\mathbb{I}(C \geq t)}{\widehat{S}^{C}_{1,1}(t|X)} 
+ \widehat{S}_{1,1}(t|X) 
\int_0^t \frac{d\widehat{M}_{1,1}^{\mathrm{C}}(r|X)}{\widehat{S}_{1,1}(r|X) \widehat{S}^{\mathrm{C}}_{1,1}(r|X)} \bigg] 
\\
& + \frac{\widehat{S}_{1,1}(t|X)}{\widehat{p}_1 - \widehat{p}_0} \bigg[\frac{1-Z}{1-\widehat{e}(X)}( D - \widehat{p}_0(X))
+ \frac{\widehat{p}_0(X)}{\widehat{p}_1(X)} \frac{Z}{\widehat{e}(X)}  (\widehat{p}_1(X) - D)\bigg]
 + \frac{\widehat{\pi}_{\co}(X)}{\widehat{p}_1 - \widehat{p}_0} 
\bigg( 1 - \frac{Z}{\widehat{e}(X)} \bigg) 
\widehat{S}_{1,1}(t|X)
\Bigg\},
\end{align}  
}
{\small
\begin{align}
    \widehat{S}^{\mathrm{mr}}_{0,\nt}(t) & = \mP_n \Bigg\{
\frac{\widehat{\pi}_{\nt}(X)}{1 - \widehat{p}_1}
\frac{1-D}{1-\widehat{p}_0(X)} 
\frac{1-Z}{ 1-\widehat{e}(X)}
\bigg[ \frac{\mathbb{I}(C \geq t)}{\widehat{S}^{C}_{0,0}(t|X) } 
+ \widehat{S}_{0,0}(t|X) 
\int_0^t \frac{d\widehat{M}_{0,0}^{\mathrm{C}}(r|X)}{\widehat{S}_{0,0}(r|X) \widehat{S}^{\mathrm{C}}_{0,0}(r|X)} \bigg] 
\\
& + \frac{\widehat{S}_{0,0}(t|X)}{1 - \widehat{p}_1} \bigg[\frac{Z}{\widehat{e}(X)}(  \widehat{p}_1(X)-D)
+ \frac{1-\widehat{p}_0(X)}{1-\widehat{p}_1(X)} \frac{1-Z}{1-\widehat{e}(X)}  (D - \widehat{p}_0(X) )\bigg]
 + \frac{\widehat{\pi}_{\co}(X)}{1 - \widehat{p}_1} 
\bigg( 1 - \frac{1-Z}{1-\widehat{e}(X)} \bigg) 
\widehat{S}_{0,0}(t|X)
\Bigg\},
\end{align}
}
{\small
\begin{align}
    \widehat{S}^{\mathrm{mr}}_{1,\nt}(t) & = \mP_n \Bigg\{
\frac{1-D}{1-\widehat{p}_1(X)} 
\frac{Z}{ \widehat{e}(X)}
\bigg[ \frac{\mathbb{I}(C \geq t)}{\widehat{S}^{C}_{1,0}(t|X) } 
+ \widehat{S}_{1,0}(t|X) 
\int_0^t \frac{d\widehat{M}_{1,0}^{\mathrm{C}}(r|X)}{\widehat{S}_{1,0}(r|X) \widehat{S}^{\mathrm{C}}_{1,0}(r|X)} \bigg] 
+ \frac{1-\widehat{p}_1(X)}{1-\widehat{p}_1} \bigg(1-\frac{Z}{\widehat{e}(X)} \bigg) \widehat{S}_{1,0}(t|X)
\Bigg\},
\end{align}
}
{\small
\begin{align}
    \widehat{S}^{\mathrm{mr}}_{0,\at}(t) & = \mP_n \Bigg\{
\frac{D}{\widehat{p}_1(X)} 
\frac{1-Z}{1- \widehat{e}(X)}
\bigg[ \frac{\mathbb{I}(C \geq t)}{\widehat{S}^{C}_{0,1}(t|X) } 
+ \widehat{S}_{0,1}(t|X) 
\int_0^t \frac{d\widehat{M}_{0,1}^{\mathrm{C}}(r|X)}{\widehat{S}_{0,1}(r|X) \widehat{S}^{\mathrm{C}}_{0,1}(r|X)} \bigg] 
+ \frac{\widehat{p}_0(X)}{\widehat{p}_0} \bigg(1-\frac{1-Z}{1-\widehat{e}(X)} \bigg) \widehat{S}_{0,1}(t|X)
\Bigg\},
\end{align}
}
{\small
\begin{align}
    \widehat{S}^{\mathrm{mr}}_{1,\at}(t) & = \mP_n \Bigg\{
\frac{\widehat{\pi}_{\at}(X)}{\widehat{p}_0}
\frac{D}{\widehat{p}_1(X)} 
\frac{Z}{ \widehat{e}(X)}
\bigg[ \frac{\mathbb{I}(C \geq t)}{\widehat{S}^{C}_{1,1}(t|X)} 
+ \widehat{S}_{1,1}(t|X) 
\int_0^t \frac{d\widehat{M}_{1,1}^{\mathrm{C}}(r|X)}{\widehat{S}_{1,1}(r|X) \widehat{S}^{\mathrm{C}}_{1,1}(r|X)} \bigg] 
\\
& + \frac{\widehat{S}_{1,1}(t|X)}{\widehat{p}_0} \bigg[\frac{1-Z}{1-\widehat{e}(X)}( D - \widehat{p}_0(X))
+ \frac{\widehat{p}_0(X)}{\widehat{p}_1(X)} \frac{Z}{\widehat{e}(X)}  (\widehat{p}_1(X) - D)\bigg]
 + \frac{\widehat{\pi}_{\at}(X)}{\widehat{p}_0} 
\bigg( 1 - \frac{Z}{\widehat{e}(X)} \bigg) 
\widehat{S}_{1,1}(t|X)
\Bigg\},
\end{align} 
}
where $
d\widehat{M}_{z,d}^C(t \mid X) = dN^C(t) - \mathbb{I}(C \geq t) \frac{\widehat{h}_{z,d}^C(t \mid X)}{\widehat{S}_{z,d}^C(t \mid X)}.
$

\textbf{Weighted standardized
mean differences}
 \begin{align}
    \text{SMD}_{\co} & = \frac{1}{\sigma_{\co}}\bigg[\mP_n \big[\frac{Z D W_{1,\co}(X) X}{\mP_n[ZD]} \big]
    - \mP_n \big[\frac{(1-Z) (1-D) W_{0,\co}(X) X}{\mP_n[(1-Z) (1-D)]} \big] \bigg] \\
     \text{SMD}_{\nt} & = \frac{1}{\sigma_{\nt}}\bigg[\mP_n \big[\frac{Z(1-D) W_{1,\nt}(X) X}{\mP_n[Z(1-D)]} \big]
    - \mP_n \big[\frac{(1-Z) (1-D) W_{0,\nt}(X) X}{\mP_n[(1-Z) (1-D)]} \big] \bigg] \\
      \text{SMD}_{\at} & = \frac{1}{\sigma_{\at}}\bigg[\mP_n \big[\frac{ZD W_{1,\nt}(X) X}{\mP_n[ZD]} \big]
    - \mP_n \big[\frac{(1-Z)DW_{0,\at}(X) X}{\mP_n[(1-Z)D]} \big] \bigg],
 \end{align}
where $W_{z,u}(X)$ is the weight, $\sigma_{\co} = \sqrt{(s^2_{11} +s^2_{00} )/2}$, $\sigma_{\nt} = \sqrt{(s^2_{10} +s^2_{00} )/2}$, $\sigma_{\at} = \sqrt{(s^2_{11} +s^2_{01} )/2}$,  and $s^2_{zd}$ is the empirical variance  of $X$ within the subjects in the $(Z=z, D=d)$ cell. 
When $W_{z,u}(X) = 1$, the SMDs measure the
systematic difference across different $(Z,D)$ cell, and therefore reflects the amount of confounding due to the ICE. When $W_{z,u}(X)$ is specified as the corresponding principal
score weights, that is,
$W_{1,\co}(X) = \frac{\widehat{\pi}_{\co}(X)}{\widehat{p}_1(X)}/\frac{\widehat{\pi}_{\co}}{\widehat{p}_1}$, 
$W_{0,\co}(X) = \frac{\widehat{\pi}_{\co}(X)}{1-\widehat{p}_0(X)}/\frac{\widehat{\pi}_{\co}}{1-\widehat{p}_0}$, 
$W_{1,\nt}(X)=1$,
$W_{0,\nt}(X) = \frac{\widehat{\pi}_{\nt}(X)}{1-\widehat{p}_0(X)}/\frac{\widehat{\pi}_{\nt}}{1-\widehat{p}_0}$, 
$W_{1,\at}(X) = \frac{\widehat{\pi}_{\at}(X)}{\widehat{p}_1(X)}/\frac{\widehat{\pi}_{\at}}{\widehat{p}_1}$, 
$W_{0,\at}(X) = 1$
, the SMDs quantifies the extent to which the confounding is addressed by weighting.


\subsection*{C. Additional results of the case study}

\begin{table}
\centering
\setlength\tabcolsep{6pt} 
\caption{Composition of $(Z,D)$ with different threshold $d$  in the CALGB 90206 Trial. } \label{tab:comp-real}
\resizebox{\textwidth}{!}{%
\begin{tabular}{c cccccc}
    \toprule
    & \multicolumn{2}{c}{$d=60$} &  \multicolumn{2}{
        c}{$d=90$} & \multicolumn{2}{
        c}{$d=120$} \\
    \cmidrule(lr){2-3} \cmidrule(lr){4-5} \cmidrule(lr){6-7}
    & \textbf{Continued} & \textbf{Discontinued} & \textbf{Continued} & \textbf{Discontinued} & \textbf{Continued} & \textbf{Discontinued} \\
    \textbf{Treatment} & ($D=0$) & ($D=1$) & ($D=0$) & ($D=1$) & 
 ($D=0$) & ($D=1$) \\  
    \midrule  
    \textbf{IFN} ($Z = 0$) & 242 & 121 & 164  & 199 & 138 &225\\
    \textbf{Bevacizumab plus IFN} ($Z = 1$) & 273   & 96 & 214  & 155 & 202 & 167 \\
    \bottomrule
\end{tabular}
}
\end{table}

\begin{table}[ht]
\centering
\caption{Stratum-specific summary of covariates in the CALGB 90206 trial using the Bayesian mixture model under different combinations of ER and monotonicity assumption}
\resizebox{\textwidth}{!}{%
\begin{tabular}{lcccccccc}
\toprule 
\textbf{Variable} 
& \multicolumn{2}{c}{$\mathbf{U=(0,0)}$} & \multicolumn{2}{c}{$\mathbf{U=(1,0)}$} & \multicolumn{2}{c}{$\mathbf{U=(1,1)}$} & \multicolumn{2}{c}{$\mathbf{U=(0,1)}$}\\
 & \textbf{Mean} & \textbf{95\% CI} & \textbf{Mean} & \textbf{95\% CI} & \textbf{Mean} & \textbf{95\% CI}& \textbf{Mean} & \textbf{95\% CI} \\
\midrule
\multicolumn{9}{c}{Monotonicity, no ER} \\
\midrule
Male ($\%$) & 69.8 & (65.1, 72.9)  &  65.9 & (30.5, 88.3) & 69.0 & (65.5, 74.1) &--&--  \\
Nephrectomy ($\%$) & 12.0 & (9.1, 14.7) & 14.9 & (3.3, 40.6) & 19.0 & (15.7, 21.7) &  --&-- \\
0 risk factors ($\%$) & 27.6 & (26.8, 28.4) & 27.1 & (20.5, 32.1)   & 24.7 & (23.8, 25.5)  &  --&-- \\
1-2 risk factors ($\%$) & 66.1 & (64.5, 67.6) & 63.6 & (51.0, 69.9)   & 60.6 & (58.9, 62.3)  &  --&-- \\
$\geq 3$ risk factors ($\%$)& 6.3 &(4.3, 8.4)  & 9.3  &(1.5, 27.6)  &14.7  &(12.3, 17.1)   &--  & -- \\
Age (years) & 61.2  &(60.3,  62.1) & 63.8  &(57.5,  70.0) &   62.5  &(61.5,  63.5) & -- & --\\
\midrule
\multicolumn{9}{c}{Monotonicity, ER} \\
\midrule
Male ($\%$) & 69.4 & (65.9, 73.0)  &  57.9 & (21.5, 84.2) & 70.3& (66.4, 74.0) &--&--  \\
Nephrectomy ($\%$) & 11.9 & (8.9, 14.9) & 17.3 & (3.0, 34.8) & 18.7 & (15.7, 22.1) &  --&-- \\
0 risk factors ($\%$) & 27.6 & (26.8, 28.5) & 27.1 & (21.5, 31.5)   & 24.6 & (23.7, 25.6)  &  --&-- \\
1-2 risk factors ($\%$) & 66.1 & (64.4, 67.7) & 63.9 & (53.5, 69.7)   & 60.6 & (58.8, 62.5)  &  --&-- \\
$\geq 3$ risk factors ($\%$)& 6.3 &(4.0, 8.6)  & 9.0  &(1.6, 23.6)  &14.8  &(12.1, 17.3)   &--  & -- \\
Age (years) & 61.2  &(60.4,  62.0) & 64.4  &(55.5,  70.1) &   62.5  &(61.6,  63.5) & -- & --\\
\midrule
\multicolumn{9}{c}{No monotonicity, no ER } \\
\midrule
Male ($\%$) & 70.1 & (65.6, 74.5)  &  66.3 & (44.7, 84.1) & 70.0 & (64.4, 75.3) & 57.2& (18.1, 82.2) \\
Nephrectomy ($\%$) & 12.3  & (9.0, 15.3) & 11.6 & (2.6, 24.5) & 20.6  & (16.8, 25.0) &  8.2 & (1.4,  23.8) \\
0 risk factors ($\%$) & 27.5 & (26.5, 28.3) & 28.5 & (24.7, 31.5)   & 24.0 & (22.5, 25.1)  &  29.7 & (24.9, 33.7) \\
1-2 risk factors ($\%$) & 66.0 & (64.4, 67.6) & 66.4 & (60.5, 70.2)   & 59.6 & (57.0, 62.0)  &  64.9 &(57.0, 69.1) \\
$\geq 3$ risk factors ($\%$)& 6.5  &(4.3, 8.9)  & 5.1  &(0.7, 14.0)  & 16.4  &(13.1, 20.3)   & 5.4 &(0.9,  17.0)\\
Age (years) & 61.1  &(60.1,  62.1) & 63.1  &(58.9,  67.5) &   62.3  &(61.1,  63.6) & 64.3 & (56.0, 70.7)\\
\midrule
\multicolumn{9}{c}{No monotonicity, ER } \\
\midrule
Male ($\%$) & 70.2 & (65.6, 74.4)  &  64.8 & (44.1, 81.7) & 70.0 & (64.9, 75.2) & 57.8& (19.7, 81.6) \\
Nephrectomy ($\%$) & 12.4  & (9.4, 15.5) & 11.6 & (2.1, 27.0) & 20.5 & (16.3, 24.9) &  8.6 & (1.6,  24.6) \\
0 risk factors ($\%$) & 27.4 & (26.5, 28.3) & 28.7 & (24.7, 31.6)   & 23.8& (22.4, 25.2)  &  29.5 & (25.0, 33.3) \\
1-2 risk factors ($\%$) & 66.0 & (64.2, 67.5) & 67.0 & (61.5, 70.3)   & 59.4 & (56.6, 61.9)  &  65.0 &(56.9, 69.1) \\
$\geq 3$ risk factors ($\%$)& 6.6  &(4.6, 8.9)  & 4.4  &(0.7, 12.8)  & 16.7  &(13.1, 20.6)   & 5.5 &(0.9,  16.6)\\
Age (years) & 61.2  &(60.3,  62.1) & 62.9  &(58.2,  68.6) &   62.5  &(61.4,  63.7) & 64.0 & (56.2, 69.7)\\
\bottomrule
\end{tabular}
\label{tab:stra-X-mixture-SA}
}
\end{table}

\begin{table}[ht]
\centering
\caption{Stratum-specific summary of pretreatment variables in the CALGB 90206 trial using the weighting method under principal ignorability and monotonicity}
\begin{tabular}{lcccccc}
\toprule
\textbf{Variable} 
& \multicolumn{2}{c}{$\mathbf{U=(0,0)}$} & \multicolumn{2}{c}{$\mathbf{U=(1,0)}$} & \multicolumn{2}{c}{$\mathbf{U=(1,1)}$} \\
 & \textbf{Mean} & \textbf{SD} & \textbf{Mean} & \textbf{SD} & \textbf{Mean} & \textbf{SD} \\
\midrule
Male ($\%$)&69.4  &46.0 &70.2  &46.5  & 69.4 &46.0   \\
Nephrectomy ($\%$)& 15.4 &36.1 &15.7&36.5 & 15.4 &36.1 \\
0 risk factors ($\%$)& 26.6 &44.2 &26.2 &44.1 & 26.6 &44.2    \\
1-2 risk factors ($\%$)& 62.7 &48.3 & 65.4 &48.2 & 62.7 &48.3   \\
$\geq 3$ risk factors ($\%$)&10.5 &30.7 & 10.0 &29.7 & 10.5& 30.7 \\
Age (years) & 61.9 &10.0& 62.9 &10.1& 61.9 & 10.0  \\
\bottomrule
\end{tabular}
\label{tab:stra-X-weighting}
\end{table}

\newpage
\begin{figure}[h]
  \centering
    \includegraphics[width=0.9\linewidth]{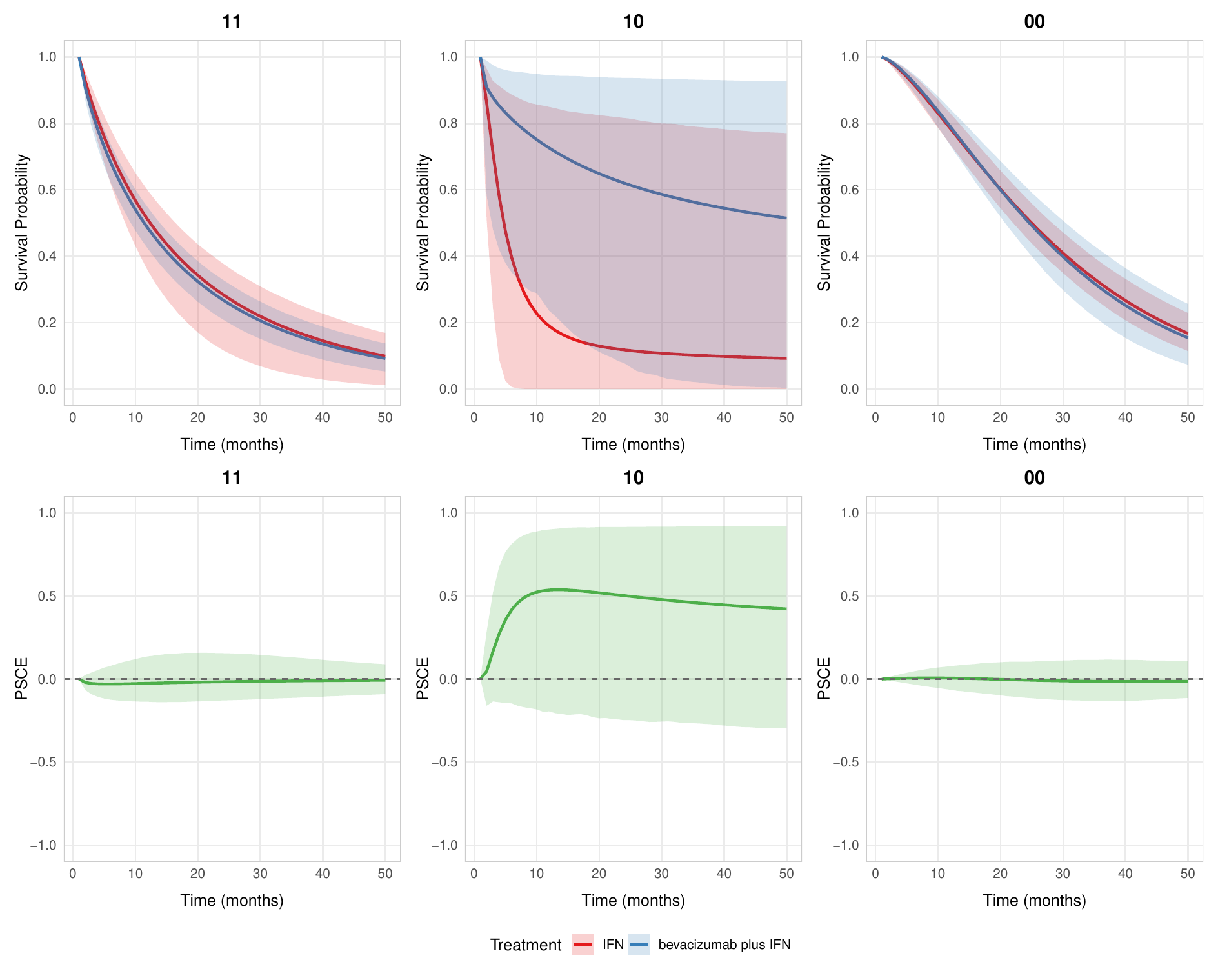} 
    \caption{Posterior survival probability curves and treatment effects of each principal stratum in the CALGB 90206 trial, obtained via the Bayesian mixture model method with the monotonicity assumption but without the ER assumption.}
   \label{fig:SPCE-Mix-Mono}
\end{figure}

\newpage
 \begin{figure}
  \centering
    \includegraphics[width=0.9\linewidth]{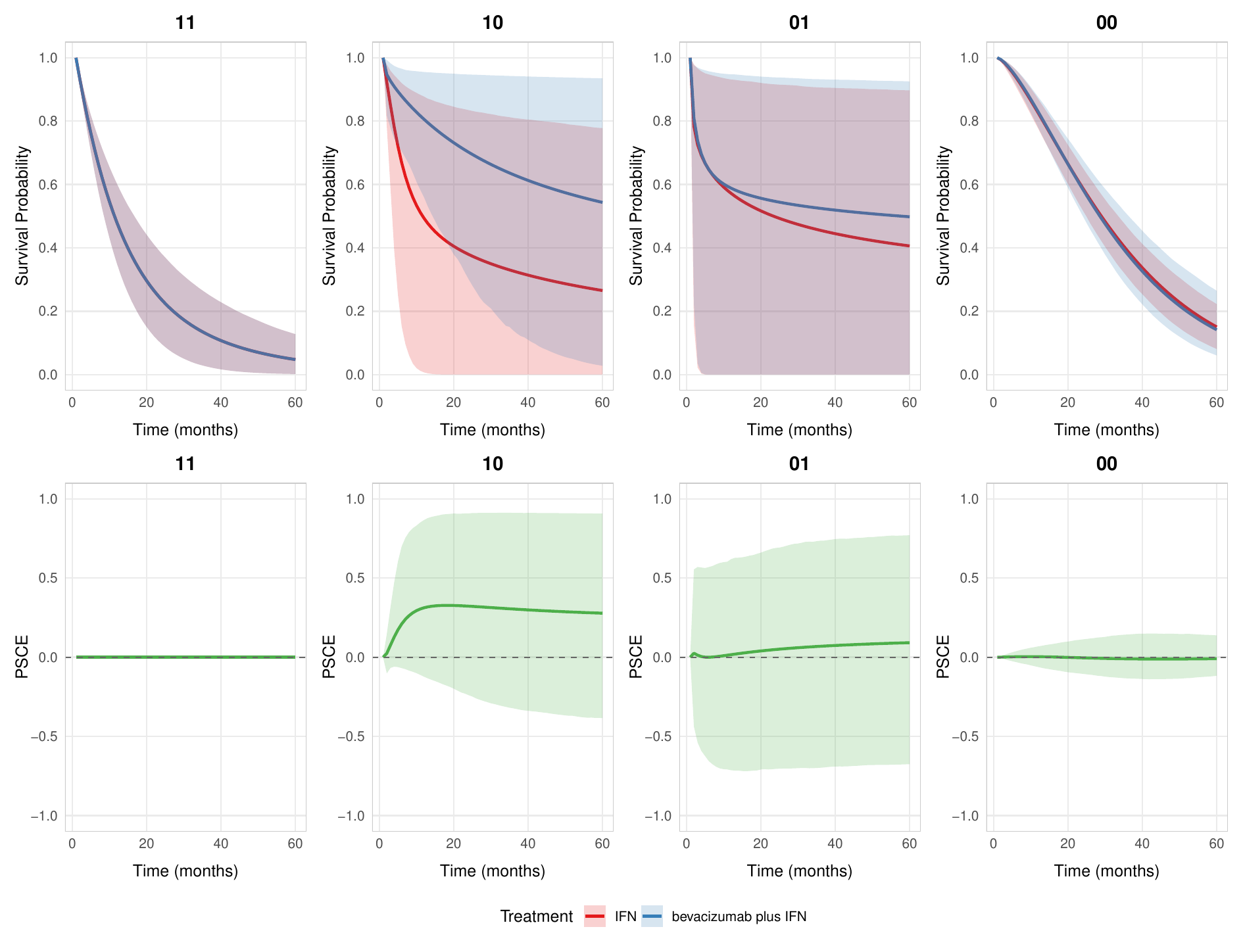} 
    \caption{Posterior survival probability curves and treatment effects of each principal stratum in the CALGB 90206 trial, obtained via the Bayesian mixture model method without the monotonicity assumption but with the ER assumption.}
   \label{fig:SPCE-Mix-ER}
\end{figure}

\clearpage
\newpage
 \begin{figure}
  \centering
    \includegraphics[width=0.9\linewidth]{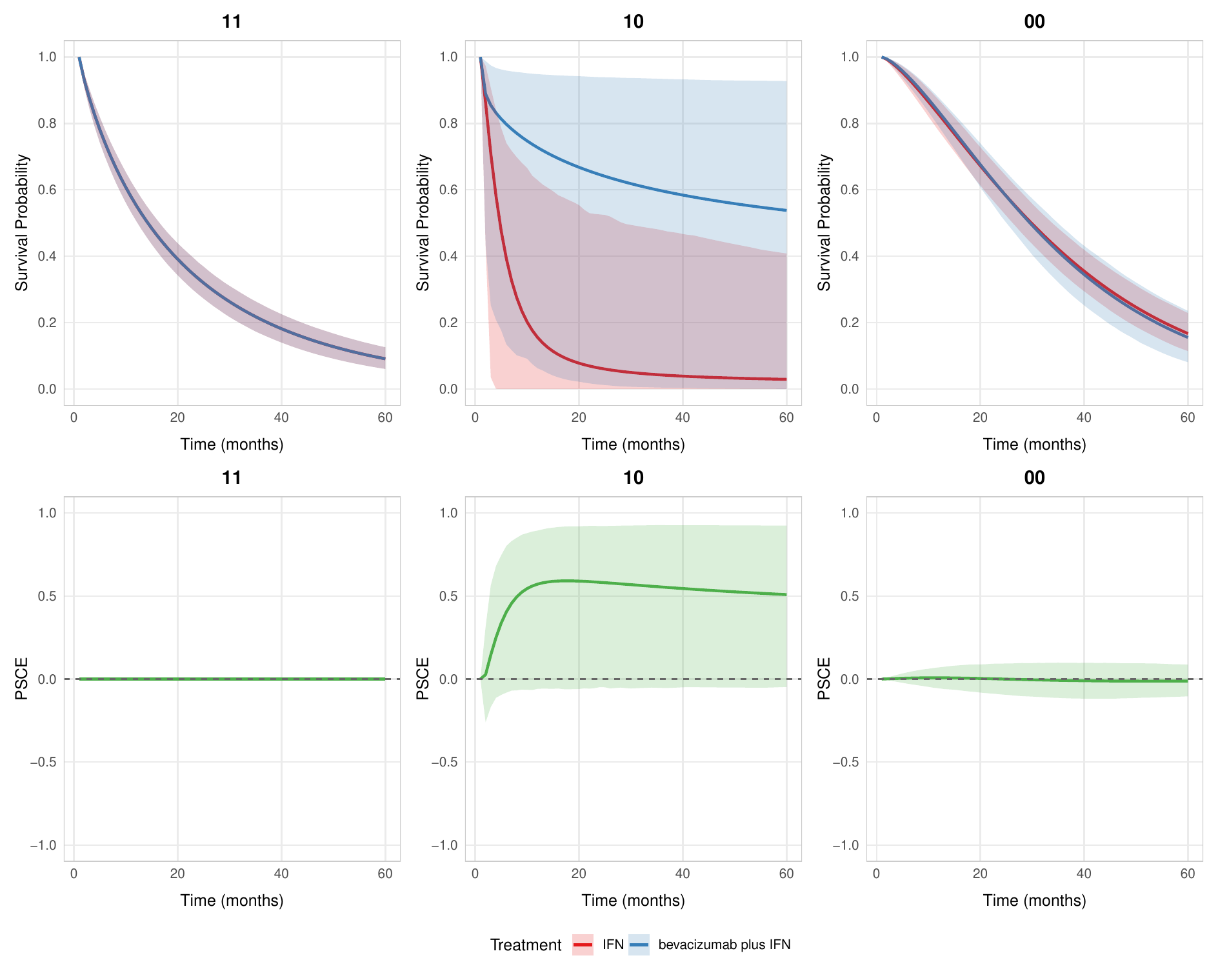} 
    \caption{Posterior survival probability curves and treatment effects of each principal stratum in the CALGB 90206 trial, obtained via the Bayesian mixture model method with both the monotonicity the ER assumption.}
   \label{fig:SPCE-Mix-Mono-ER}
\end{figure}

\clearpage

\begin{figure}
    \centering
    \includegraphics[width=0.75\linewidth]
      {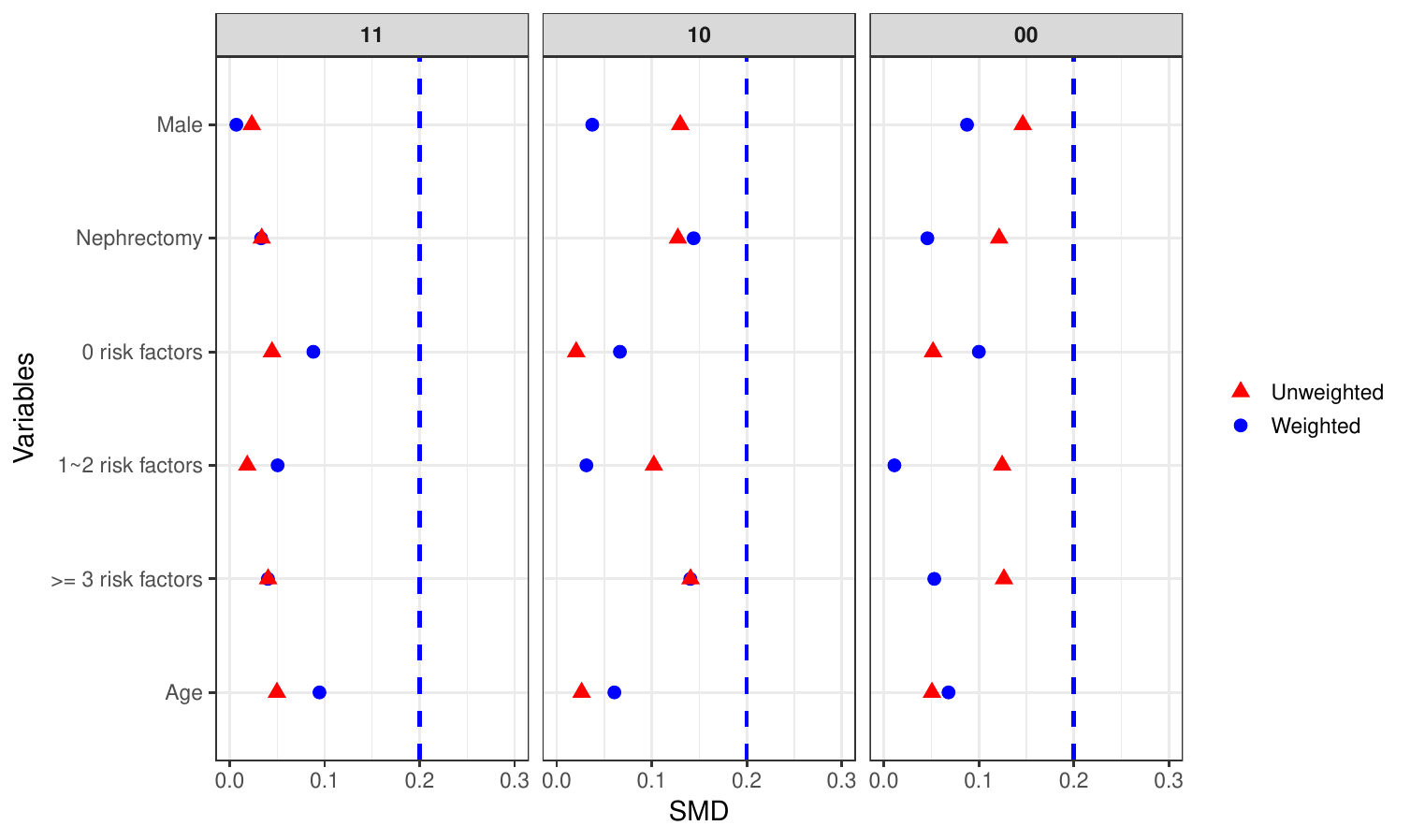} 
    \caption{ Weighted standardized
mean differences (SMD) of baseline covariates among strata in the CALGB 90206 trial.} \label{fig:SMD}
\end{figure}

\clearpage
\newpage
\begin{figure}
    \centering
    \includegraphics[width=0.9\linewidth]{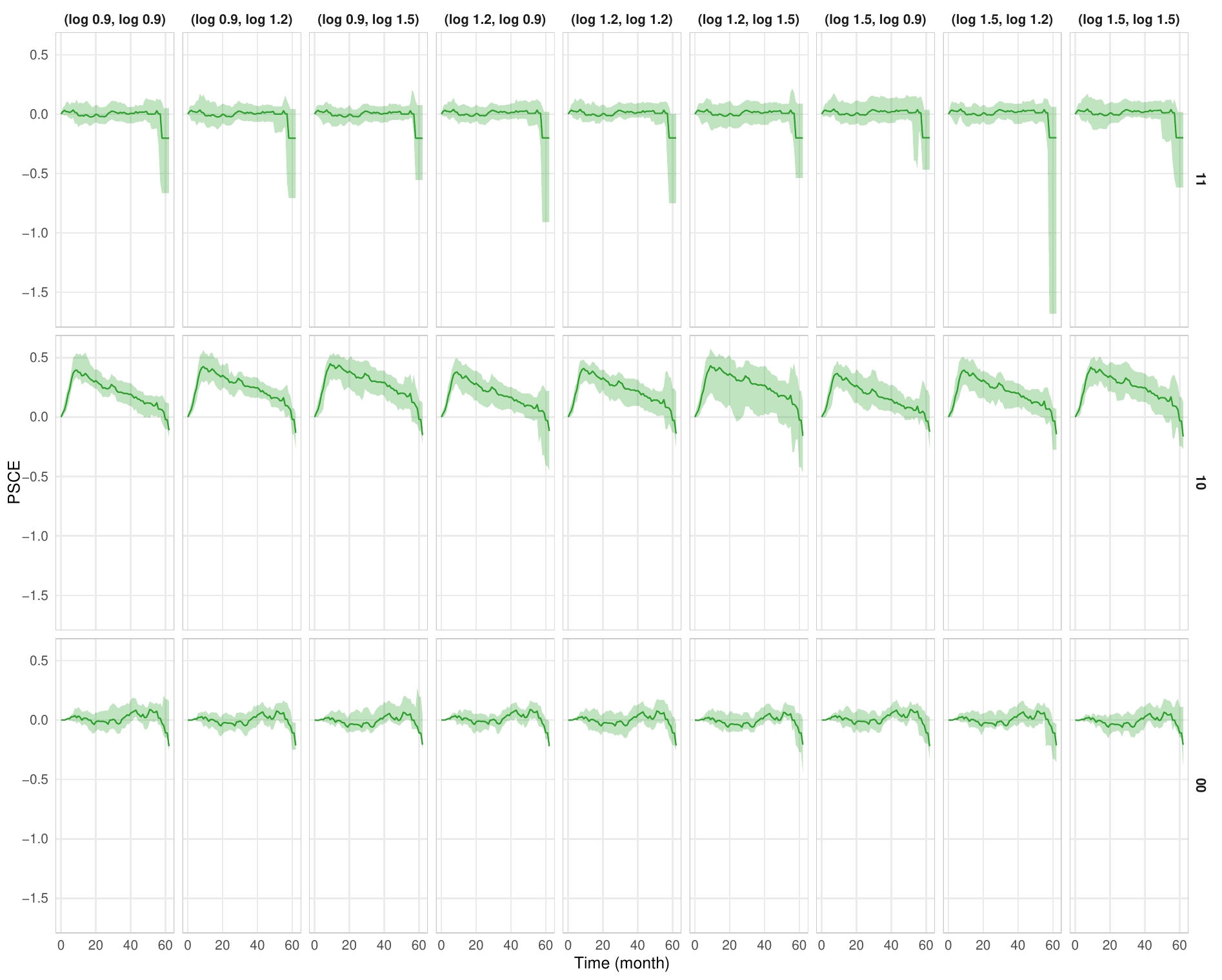} 
    \caption{The PSCE estimates in the sensitivity analysis of the principal ignorability assumption with different $(\eta_0, \eta_1) $ values under the weighting method in the CALGB 90206 trial. } \label{fig:SPCE-PI-sen}
\end{figure}

\clearpage
\newpage
\subsection*{D. Simulation Detail}
\begin{table}[htbp]
\centering
\caption{Coefficients $\phi_{z, u}$, $\psi_{z,u}$ and $\gamma_{z, u}$ to generate outcomes in the simulation.}
\begin{tabular}{ccccc}
\toprule
$u$ & $z$ & $\phi_{z, u}$ & $\psi_{z, u}$ & $\gamma_{z, u}$ \\
\midrule
(1, 1) & 0 & 2.0 & -5.0 & (-0.3, 0.3, 1.0, 1.0) \\
(1, 1) & 1 & 2.0 & -5.2 & (-0.3, 0.3, 1.0, 1.0) \\
(0, 1) & 0 & 2.0 & -4.8 & (-0.3, 0.3, 1.0, 1.0) \\
(0, 1) & 1 & 1.6 & -5.5 & (-0.5, 0.6, 1.5, 1.5) \\
(1, 0) & 0 & 2.5 & -4.0 & (-0.5, 0.6, 0.5, 1.0) \\
(1, 0) & 1 & 2.5 & -4.5 & (-0.5, 0.6, 0.5, 0.5) \\
(0, 0) & 0 & 2.2 & -4.5 & (-0.6, 0.5, 0.9, 0.5) \\
(0, 0) & 1 & 1.6 & -5.5 & (-0.5, 0.6, 1.4, 1.5) \\
\bottomrule
\end{tabular}\label{tab:sim_Y}
\end{table}

\end{document}